\documentclass[10pt,journal]{IEEEtran}
\usepackage{color}
\usepackage{cite,amsmath}
\usepackage[numbers,sort&compress]{natbib}
\usepackage{graphicx}
\usepackage{epstopdf}
\usepackage{caption}
\usepackage{float}
\usepackage{subfigure}
\usepackage[justification=centering]{caption}
\usepackage{amsfonts,amssymb}
\usepackage{multirow}
\usepackage{algorithm}
\usepackage{algorithmicx}
\usepackage{algpseudocode}
\usepackage{amsmath}
\usepackage{diagbox}
\usepackage{cite}
\usepackage{booktabs}

\ifCLASSINFOpdf
\else

\fi

\begin{document}

\title{Variational Bayesian Learning based Joint Localization and Path Loss Exponent with Distance-dependent Noise in Wireless Sensor Network}

\author{Yunfei Li, Yiting Luo, Weiqiang Tan, Chunguo Li, Shaodan Ma and Guanghua Yang
\thanks{Y. Li  and Yiting Luo are with the Department of Electrical Engineering, Anhui Polytechnic University, Wuhu, China (email: $\rm lyf@mail.ahpu.edu.cn$, $\rm lyt1222@ahpu.edu.cn$).}
\thanks{W. Tan is with the School of Computer Science and Cyber Engineering, Guangzhou University,
Guangzhou, China  (e-mail: $\rm wqtan@gzhu.edu.cn$).}
\thanks{C. Li is with the School of Information Science and Engineering, Southeast University, Nanjing, China
Purple Mountain Laboratories, Nanjing, China (e-mail: $\rm chunguoli@seu.edu.cn$).}
\thanks{S. Ma is with the State Key Laboratory of Internet of Things for Smart City and the Department
of Electrical and Computer Engineering, University of Macau, Taipa, Macao, China (e-mail: $\rm shaodanma@um.edu.mo$).}
\thanks{G. Yang is with the School of Intelligent Systems Science and Engineering, Jinan University, Zhuhai 519070, China (e-mails:ghyang@jnu.edu.cn).}
\vspace{-9mm}
}

\maketitle

\begin{abstract}
This paper focuses on the challenge of jointly optimizing location and path loss exponent (PLE) in distance-dependent noise. Departing from the conventional independent noise model used in localization and path loss exponent estimation problems, we consider a more realistic model incorporating distance-dependent noise variance, as revealed in recent theoretical analyses and experimental results. The distance-dependent noise introduces a complex noise model with unknown noise power and PLE, resulting in an exceptionally challenging non-convex and nonlinear optimization problem. In this study, we address a joint localization and path loss exponent estimation problem encompassing distance-dependent noise, unknown parameters, and uncertainties in sensor node locations. To surmount the intractable nonlinear and non-convex objective function inherent in the problem, we introduce a variational Bayesian learning-based framework that enables the joint optimization of localization, path loss exponent, and reference noise parameters by leveraging an effective approximation to the true posterior distribution. Furthermore, the proposed joint learning algorithm provides an iterative closed-form solution and exhibits superior performance in terms of computational complexity compared to existing algorithms. Computer simulation results demonstrate that the proposed algorithm approaches the performance of the Bayesian Cramer-Rao bound (BCRB), achieves localization performance comparable to the (maximum likelihood-Gaussian message passing) ML-GMP algorithm in some cases, and outperforms the other comparison algorithm in all cases.
\end{abstract}

\begin{IEEEkeywords}
 BCRB, parameter estimation, distance-dependent, localization and sensor node uncertainty
 \vspace{-2em}
\end{IEEEkeywords}

\IEEEpeerreviewmaketitle

\section{Introduction}
\subsection{Motivation and Literature Review}
Within the scope of location optimization, the tasks of target localization hold paramount importance in various communication services. These services encompass critical aspects such as resource allocation, beamforming, and synchronization, playing key roles in wireless applications, which include equipment tracking in Industrial Internet of Things (IIoTs) \cite{ApproachMekala23IoT,23IoTJointSale}, tracking of autonomous vessels in Ocean of Things (OoTs) \cite{HuoCellular20IoT, BaiIoTRandom20}, and staff localization in healthcare monitoring systems \cite{TVT23NazariRadio, ZhangTWCPassive23}, among others. The advent of these emerging applications has generated a pressing requirement for accurate positions, which is fundamental to addressing the intricate requirements posed by diverse wireless services, further highlighting the critical role of localization in affecting the performances of communication systems.

To tackle the complex localization challenges posed by the various emerging applications, researchers have studied and proposed advanced techniques and algorithms, such as a Newtonized variational inference spectral locolization algorithm in the millimeter-wave (mmWave) communication systems \cite{YangTSPSoft22}, a two-stage fingerprint-based localization method in the cell-free massive multiple input multiple outputs (MIMO) Internet of Things (IoT) systems \cite{WeiIoTLocalization22}, an extended space-alternating generalized expectation-maximization (SAGE) algorithm \cite{HongTWCJoint23} in multipath environment, a variational Bayesian inference localization algorithm in RIS-aided system \cite{LiTWC24Variational} and the polybolck outer approximation algorithm in time difference of arrival (TDOA) tracking \cite{LiMultiobjective19TVT}. Furthermore, a variety of models and algorithms addressing the localization problem can be found in \cite{KazazTWC22Delay, ZhouTVTEffect22, ZhengTSPJoint20, WenTVTJoint18}, presenting the diverse range of approaches developed to address the localization in different communication scenarios (For more references and specifics, please refer to \cite{ChenCSTTutorial22, LaoudiasCSTSurvey18, CSTShastriReview22}).

Among these extensive and diverse localization research endeavors, the major underlying theoretical assumption relies on distance-independent Gaussian measurement noises, which significantly reduces localization complexity. However, the experimental results and theoretical analysis both revealed that the time of arrival (TOA) and time of difference arrival (TDOA)  noises are distance-dependent under certain conditions, indicating that the noise variances are directly correlated with the unknown target-sensor distances. This complex noise model with unknown noise variance brings significant challenges to accurate target localization.

Considering the intricate nature of the localization problem with the distance-dependent noise model, existing research has predominantly focused on two primary aspects: the sensor placement problem and the target node localization problem. In addressing the sensor placement problem, a common approach involves maximizing the Fisher Information Matrix (FIM) or minimizing the Cramer-Rao Lower Bound (CRLB) through strategically placing sensors that leverage different measurements, including range, bearing, received signal strength (RSS) \cite{ConstrainedLiangGRSL16}, angle of arrival (AoA) \cite{AccessOptimalWang19}, time of arrival (ToA) \cite{FormationYanCL14}, and time difference of arrival (TDoA) \cite{HuangTWC15Source}. These endeavors aim to optimize sensor configurations to enhance the accuracy of localization outcomes. In the target node localization problem with distance-dependent noise, research efforts have been directed toward developing effective algorithms. In \cite{HuangTWC15Source}, an iterative reweighted generalized least square (IRGLS) localization algorithm was proposed for target node localization. However, to mitigate potential divergence issues and reduce computational complexity, the proposed algorithm approximated the weighted covariance matrix by neglecting correlations between its elements, which imposes limitations on its overall localization performance and convergences. Another work is presented in \cite{WCLLiRobust21}, where the target node localization problem is addressed along with consideration for sensor location uncertainties. This work introduces a two-stage maximum likelihood-Gaussian message passing algorithm (ML-GMP) that transforms distance-dependent noise into distance-independent noise leveraging the maximum likelihood estimator with a high signal-to-noise ratio (SNR). Then the GMP algorithm is employed to estimate the target node location, offering a robust solution to the challenges posed by the distance-dependent noise model. Furthermore,  research studies such as \cite{NoncooperativeIoT20Li, LiTVTCooperative22} have investigated the application of belief propagation-based localization and target tracking algorithms in mobile target node tracking with autonomous underwater vehicles (AUVs). These efforts presented the various approaches and methodologies employed to address challenges associated with distance-dependent noise in the localization and tracking of mobile targets.

The research works discussed above primarily focus on either the sensor placement problem or the target node localization problem, neglecting the critical aspect of environment-varying path loss exponent, unknown reference noise power, and inaccurate sensor node locations. In practical scenarios, obtaining perfect prior information, including parameters such as the path loss exponent and reference power noise, is challenging and even unattainable. Furthermore, achieving accurate sensor location awareness, especially with mobile sensors like autonomous underwater vehicles or unmanned aerial vehicles (UAVs), is inherently difficult due to their mobility and the inevitability of location errors \cite{TCRobustLi2020, LiSecure20IoT}. Consequently, a more comprehensive approach is required, which addresses the challenges of joint estimation, imperfect sensor location awareness, and the intricacies of distance-dependent noise, and provides realistic and robust solutions for the joint estimation problem in the distance-dependent model.

\subsection{Contributions}

In this paper, our primary focus is addressing the intricate challenge posed by the joint estimation of location, PLE, and reference noise parameters, specifically in distance-dependent noise environments where sensor location errors are present. The complexity arises from the fact that the measurement noise is determined by the interplay of the unknown path loss exponent parameter, reference power noise, and target-sensor distances. Moreover, the path loss exponent exhibits an exponential correlation with the unknown target-sensor distances, further complicated by imperfectly known sensor locations. The joint localization and parameter estimation problem in this problem is greatly challenging due to the inherent nonlinearity and non-convexity of the objective function. To tackle these challenges and derive optimal solutions, we propose a novel algorithm under the variational Bayesian learning framework.
This algorithm leverages a directed graphical model and effective approximations to posterior distributions, providing a robust and efficient approach to solving the joint localization and parameter estimation problem. In summary, the key contributions of this paper can be outlined as follows: \vspace{-0.8em}
\begin{itemize}
  \item  Practical model: Unlike prior research that considered distance-dependent noise models, our work introduces a novel model of distance-dependent noise with unknown parameters and inaccurate sensor node locations, offering a more realistic representation and enhancing the model's applicability to practical scenarios involving uncertainties. This model leads to a joint estimation problem of considerable complexity. The inherent intricacies of this problem render the quest for solutions challenging, particularly when aiming for low-complexity methodologies. Although similar variational Bayesian learning-based algorithms have been applied to other localization and estimation scenarios, such as the joint estimation of PLE and target locations with RSS measurements \cite{TSP20JinBayesian}, mixed none-line-of-sight (NLOS)/line-of-sight(LOS) localization \cite{TCRobustLi2020, EMYinTSP14}, secure localization \cite{LiSecure20IoT}, unmanned aerial vehicles (UAVs) localization \cite{CLJiang21Localization} and user location tracking in massive MIMO system \cite{LianTSP19USer}, the intrinsic differences between considered models and the distance-dependent model hinges the direct application of the proposed algorithms and requires a more robust algorithm.
  \item The intricacies of the intractable and non-convex objective function drive us to forge a variational Bayesian learning-based framework for the joint localization and parameter estimation challenge. Our proposed algorithm is crafted to iteratively approximate the true posterior, guaranteeing both precision and convergence.
      This joint estimation algorithm is a testament to innovative methodologies that deftly balance theoretical rigor with complexity. It keenly acknowledges the inherent complexities resulting from the practical noise model, thereby offering a robust and efficient solution to the demanding joint localization and parameter estimation problem.
  \item Our paper provides a detailed analysis of the computational complexity and theoretical convergence of the proposed algorithm.
This algorithm is designed to iteratively approximate the complex true posterior, and it features an iterative closed-form solution, enhancing its computational efficiency. This highlights the efficiency and effectiveness of our proposed algorithm in addressing the challenges posed by joint localization and parameter estimation.
\item We derive the complicated Bayesian Cramer-Rao Bound (BCRB) with the unknown nuisance parameters and node locations. Despite the presence of prior literature discussing similar bounds, it falls short of revealing the interplay of the intricate model parameters referenced in the paper on the estimation performance. This proposed bound serves as a benchmark, characterizing the theoretical limit of achievable estimation accuracy. In our Monte Carlo simulation results,  the proposed algorithm closely aligns with the BCRB. This indicates that our algorithm approaches the optimal performance defined by the BCRB, demonstrating its efficiency in achieving nearly optimal estimation accuracy in the joint localization and parameter estimation problem.
\end{itemize}

\subsection{Organization}
{The remainder of this paper is organized as follows. In Section II, the joint localization and parameter estimation problem with sensor location errors in distance-dependent environments is formulated. Section III introduces the detailed derivations of the variational Bayesian learning framework. Then, Section IV presents theoretical derivations of the proposed joint localization and parameter estimation algorithm.
In order to make specified interpretations of the proposed algorithm, Section IV gives the analysis of the proposed algorithm. Finally, we present the simulation results in Section V and conclude the paper in Section VI.}

\section{System Model}

We consider a network consisting of one target node and $N$ sensor nodes. The unknown location of the target node is given by ${\boldsymbol{x}} = {\left[ {x,y} \right]^T}$ and the location of  sensor node $i$ is given by ${\boldsymbol{x}}_i  = {\left[ {x_i,y_i} \right]^T}$. Meanwhile, the locations of sensor nodes are not perfectly known and are modeled as ${\boldsymbol{x}}_{i}={\boldsymbol{\bar x}}_{i}+\Delta {\boldsymbol{x}}_{i}$, where $\Delta {\boldsymbol{x}}_{i}$ represents the preliminary location error. The measurement received from the $i$-th sensor node is given by
\begin{equation}\label{system}
{r_i} = \underbrace {{{\left\| {{\boldsymbol{x}} - {{\boldsymbol{x}}_i}} \right\|}_2}}_{{d_i}} + {\eta _i},
\end{equation}
where  ${\eta _i}$ is a Gaussian noise with a zero mean and an unknown distance-dependent variance\footnote{The detailed derivations of the model by leveraging the efficient estimator of TOA and received signal power logarithm fading model, which can be found in \cite{HuangTWC15Source}.} \cite{WCLLiRobust21,HuangTWC15Source}, which is described as $\mathcal{N}\left(0, \delta_{0}^{2}\left(\frac{d_{i}}{d_{0}}\right)^{\gamma}\right)$. $d_{0}$ is the reference distance and $d_0 = 1$ for ease of presentation. $\delta^2_{0}$ is the noise power at the reference distance $d_{0}$. $\gamma$ is the unknown path loss exponent.

Given the measurements from all sensors, the likelihood function can be formulated as
\begin{equation}\label{likelihood}
\begin{aligned}
p\left(\boldsymbol{r} \mid \boldsymbol{x}, \boldsymbol{x}_{1: N}, \gamma, \delta_{0}\right)&=\prod_{i=1}^{N} p\left(\boldsymbol{r}_{i} \mid \boldsymbol{x}, \boldsymbol{x}_{i}, \gamma, \delta^2_{0}\right)\\ &=\prod_{i=1}^{N} \mathcal{N}\left(r_{i}-d_{i}, \delta_{0}^{2} d_{i}^{\gamma}\right),
\end{aligned}
\end{equation}
where $\boldsymbol r=\left[r_{1}, \ldots, r_{N}\right] $ and  $ {\boldsymbol x}_{1: N}=\left[{\boldsymbol x}_{1}^{T}, \ldots, {\boldsymbol x}_{N}^{T}\right]^{T}$.
\begin{figure}
  \centering
  \includegraphics[width=3in]{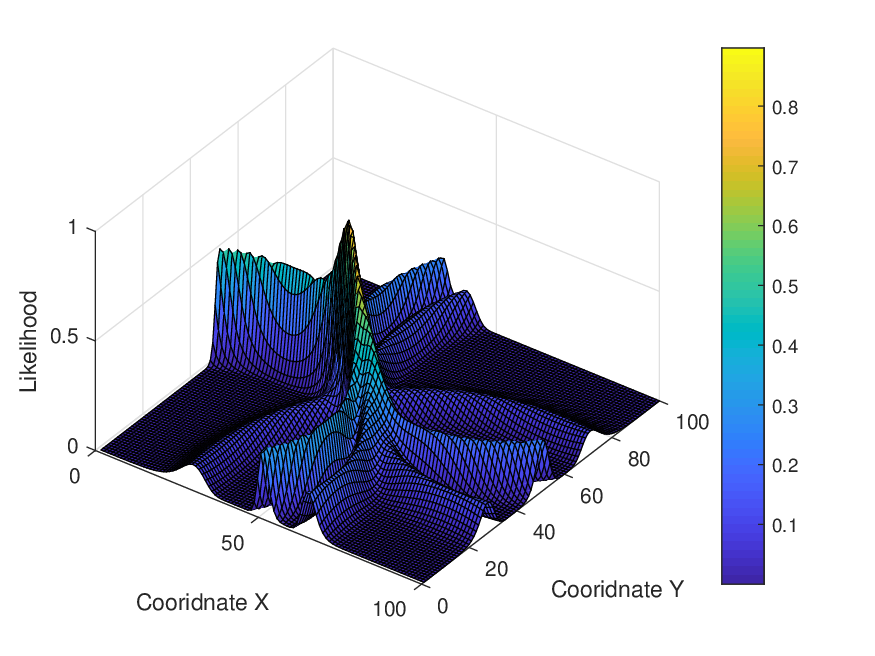}\\
  \caption{The log-likelihood distribution illustration with $N=5$, $\gamma = 4$, $\delta^2_0 = 0.001 $, ${\boldsymbol x} = {\left[ {50,50} \right]^T}$ and perfect sensor node locations}\label{fig:3d}
\end{figure}

Fig.\ref{fig:3d} compellingly unveils the intricate interplay between the target node's location and the log-likelihood function. The numerical results reveal a two-tier fact. On one hand, the precise determination of the target node's location hinges on the application of multilateralism, utilizing the sensor-target distances as depicted in Fig.\ref{fig:3d}. Nevertheless, the acquisition of accurate sensor-target distances poses a significant challenge, attributed to the presence of correlated unknown noise and the imperfect locations of sensor nodes. On the other hand, the inherent complexity of the nonlinear and multimodal log-likelihood function brings great difficulty to direct maximization with low complexity, underscoring the need for efficient approaches in addressing this intricate relationship.
In our pursuit of extracting maximum benefits from prior information, optimizing likelihood distributions, and achieving heightened accuracy in estimation solutions, we strategically embrace the foundational principle of Bayesian estimation. This approach not only empowers us to integrate existing knowledge effectively but also facilitates a comprehensive exploration of uncertainties, leading to more robust estimations. By leveraging the Bayesian framework, we tackle the intricacies of the estimation process with a posterior perspective, ensuring a thorough incorporation of available information for enhanced accuracy and reliability of our algorithm.

By considering the prior information of the unknown parameters and the Bayesian rule, the posterior distribution can be given by
\begin{equation}\label{Bayesian}
\begin{aligned}
p\left( {{\boldsymbol{\Theta }} |\boldsymbol r} \right) \propto  p\left( {{\boldsymbol{r}}\mid {\boldsymbol{\Theta }}  } \right)p\left( \gamma  \right)p\left( {\boldsymbol{x}} \right)p\left( \delta^2_{0} \right)\prod\limits_{i = 1}^N p \left( {{{\boldsymbol{x}}_i}} \right),
\end{aligned}
\end{equation}
where ${\boldsymbol{\Theta }} = \left[ {{{\boldsymbol{x}}^T},{\boldsymbol{x}}_{1:N}^T,\gamma ,{\Lambda _0}} \right]$ and ${\Lambda _{0}} = {\delta}^{-2}_0$. The estimation of the unknown parameters can be obtained via the following maximization
\begin{equation}\label{directmaximization}
{\boldsymbol{\hat \Theta }} = \arg \max p\left( {{\boldsymbol{\Theta }}|{\boldsymbol{r}}} \right).
\end{equation}

Given the measurements from all sensor nodes, the target node location and the parameters can be theoretically estimated in \eqref{directmaximization}. In the paper, we focus on the joint estimation of the target node location, the fading parameter $\gamma$, and noise reference power $\delta_{0}$ with the distance-dependent noise and sensor node location errors. The direct maximization of the likelihood function in \eqref{likelihood} and the posterior distribution in \eqref{directmaximization} are intractable due to the nonlinear distance, unknown noise variance, and coupled unknown parameters. To obtain the solutions to the joint estimation and localization problem, we proposed a variational message passing-based joint localization and parameter estimation algorithm.

\section{Variational Bayesian Learning Framework}
Based on the formulated problem in \eqref{directmaximization},  the goal of direct estimation of the unknown parameters is not tractable. Therefore, we aim to find an approximation distribution to the true posterior distribution and find the estimation of the unknown parameters. In this section, we propose a variational message passing-based joint localization and parameter estimation algorithm under the variational Bayesian learning framework.

In the Bayesian learning framework, the complicated and intractable posterior distribution motivates us to find a distribution $q({\boldsymbol{\Theta }})$ to approximate the true posterior distribution $p({\boldsymbol{\Theta }}|{\boldsymbol{r}})$. Meanwhile,  the variational distribution $q({\boldsymbol{\Theta}})$ should be tractable. By assuming the independence between variational distributions of unknown parameters and the mean-field theory, the variational distribution $q({\boldsymbol{\Theta }})$ can be factorized as
\begin{equation}\label{meanfield}
q\left( {\boldsymbol{\Theta}} \right) = \prod\limits_{{{\boldsymbol{\Theta}}_k} \in {\boldsymbol{\Theta}}} {q\left( {{{\boldsymbol{\Theta}}_k}} \right)} ,
\end{equation}
Meanwhile, the Kullback-Leibler (KL) divergence \cite{TSMVariational} is introduced to evaluate the information distance between the variational distribution $q({\boldsymbol{\Theta}})$ and the true distribution $p({\boldsymbol{\Theta }}|{\boldsymbol{R}})$, which is given by
\begin{equation}\label{KL}
\begin{aligned}
{\text{KL}}\left( {q\left( {\boldsymbol{\Theta }} \right)||p\left( {\boldsymbol{\Theta }} | {\boldsymbol r}\right)} \right)
 =  - {\mathbb{E}_{q\left( {\boldsymbol{\Theta}} \right)}}\left\{ {\ln \frac{{p\left( {\boldsymbol{\Theta}} | {\boldsymbol r}\right)}}{{q\left( {\boldsymbol{\Theta}} \right)}}} \right\} \ge 0,
 \end{aligned}
\end{equation}
where ${\mathbb{E}_{q\left( {\boldsymbol{\Theta}} \right)}}$ is the expectation with respect to ${q\left( {\boldsymbol{\Theta}} \right)}$ and the equality holds only when $q\left( {\boldsymbol{\Theta}} \right) = p\left(  {\boldsymbol{\Theta}} | {\boldsymbol r} \right)$ as shown in \cite{TCRobustLi2020}.

By substituting the mean-field factorization in \eqref{meanfield} into the Kullback-Leibler (KL) divergence in \eqref{KL} and applying an alternative optimization method
the variational distribution $q^{(\xi)}\left( {{{\boldsymbol \Theta} _k}} \right)$ can be iteratively given by \cite{TSMVariational}
\begin{equation}\label{variationalphik}
q^{(\xi)}\left( {{{\boldsymbol \Theta} _k}} \right) \propto \exp \left\{ {{\mathbb{E}_{{q^{(\xi)}\left( {{{\boldsymbol \Theta} _{\backslash k}}} \right)} }}\left[ {\ln p\left( {{\boldsymbol \Theta} ,{\boldsymbol{r}}} \right)} \right]} \right\},
\end{equation}
where $q^{(\xi)}\left( {{{\boldsymbol \Theta} _k}} \right)$ is the the $\xi$-th approximation to the true posterior distribution $p\left( {{{\bf{\Theta }}_k}|{\boldsymbol{r}}} \right)$  and $p\left( {{\boldsymbol \Theta} ,{\boldsymbol{r}}} \right)$ is the joint probability. ${\mathbb{E}_{{q^{(\xi )}}\left( {{{\boldsymbol{\Theta }}_{\backslash k}}} \right)}}$ means the expectation with respect the variational distributions excluding the variational distribution $q^{(\xi)}\left( {{{\boldsymbol \Theta} _k}} \right)$. The variational distribution in \eqref{variationalphik} will approximate the true posterior distribution iteratively and the approximated distribution ${q\left( {{{\boldsymbol{\Theta }}_k}} \right)}$ is treated as the approximation of the corresponding posterior distribution $p\left(  {\boldsymbol{\Theta}}_k | {\boldsymbol{r}} \right)$. For example, $q\left(\gamma \right) $ is the approximation to the posterior distribution $p\left( { \gamma|{\boldsymbol{r}}} \right) $.

Then the estimation of  parameter ${{\boldsymbol{\Theta }}_k}$ can be obtained via maximum a posterior(MAP) as
\begin{equation}\label{MAP_estimation}
{{\boldsymbol{\Theta }}^{{\text{MAP}}}_k}= {\mathop{\rm argmax}\nolimits} \; {q\left( {{{\boldsymbol{\Theta}}_k}} \right)}.
\end{equation}

In order to find the tractable forms of variational distribution $q\left( {{{\boldsymbol \Theta}}} \right)$, we assume the prior distributions and the variational distribution follows the conjugate prior principles, which guarantees the variational distribution $q^{(\xi)}\left( {{{\boldsymbol \Theta} _k}} \right)$ is identical to the prior distribution $p\left( {{{\boldsymbol \Theta} _k}} \right)$ in form, i.e., $p\left( \gamma \right)$ is a Gaussian distribution with ${\cal N}\left( {\gamma |{\mu _\gamma },{\delta _\gamma }} \right)$ then the $\xi$-th variational distribution
$q^{(\xi)}\left(\gamma \right)$ is also a Gaussian distribution with $\mathcal{N}\left( {\gamma |{\mu^{\left( \xi  \right)} _\gamma },\delta _\gamma ^{\left( \xi  \right)}} \right)$ and other prior distributions are clarified as following:
\begin{itemize}
  \item For the reference noise power $\delta_0$, the theoretical results signify that its inverse variable $\Lambda_0$ can be modeled as a gamma distribution and is given by
  \begin{equation}\label{gammadis}
  p\left( {{\Lambda _0}} \right) = \Gamma \left( {{\Lambda _0}|a,b} \right).
  \end{equation}
  \item The sensor node locations are assumed to be  known with location uncertainty and are modeled as
  \begin{equation}\label{sensoruncertain}
  p\left( {{{\boldsymbol x}_i}} \right) = \mathcal{N}\left( {{{\boldsymbol x}_i}|{{\bar {\boldsymbol x}}_i},{\boldsymbol \Sigma _i}} \right),
  \end{equation}
  where ${{\bar {\boldsymbol x}}_i}$ is the known coarse location and ${\boldsymbol \Sigma _i}$ is the covariance matrix.
  \item  The prior distribution of the target node location is also assumed to be a Gaussian distribution with $p\left( {{{\boldsymbol x}}} \right) = \mathcal{N}\left( {{{\boldsymbol x}}|{{\bar {\boldsymbol x}}},{\boldsymbol \Sigma }} \right)$ and ${{\bar {\boldsymbol x}}}$ is the known coarse location and ${\boldsymbol \Sigma}$ is the covariance matrix.
\end{itemize}

Considering the vast complexity of repetitive and boring computation with the joint distribution $p\left( {{\boldsymbol \Theta},{\boldsymbol{r}}} \right)$, we further build a probabilistic graphical model and map the dependencies between the variables in the directed graph. As shown in Fig.1,  the variables are represented by nodes in the directed graph, and the nodes are connected by the directed edges, which are the conditional dependencies between the variables.
Moreover, the nodes can be classified into parent nodes ${\cal P}_{{\boldsymbol \Theta _k}}$ and child nodes ${{\cal C}_{{\boldsymbol \Theta _k}}}$ for a variable node ${\boldsymbol \Theta _k}$  according to Bayeisan graphical theory. For example,  the node $a$ and node $b$ from the prior distribution \eqref{gammadis} are the parent nodes ${\cal P}_{\Lambda _0}$ and node $\boldsymbol r$ from the likelihood function is the child node ${\cal C}_{\Lambda _0}$ of parameter ${\Lambda _0}$ respectively.
Using the Bayeisan graphical representation, the marginal probability $p\left( {{\boldsymbol \Theta _k}|{{\cal P}_{{\boldsymbol \Theta _k}}}} \right)$ and the conditional dependencies $p\left( {{{\cal C}_{{\boldsymbol \Theta _k}}}|{\boldsymbol \Theta _k},{\boldsymbol \Theta _m}} \right)$, joint probability distribution $p\left( {{\boldsymbol \Theta} ,{\boldsymbol{r}}} \right)$ can be reformulated.

Invoked by the fact that the prior distributions and likelihood distribution in the probabilistic graphical model all follow the form of exponential functions, we can reformulate the conditional distributions as
\begin{equation}\label{parentdis}
p\left( {{\boldsymbol \Theta _k}|{{\cal P}_{{\boldsymbol \Theta _k}}}} \right) \propto \exp \left[ {{{\cal J}_{{\boldsymbol \Theta _k}}}\left( {{{\cal P}_{{\boldsymbol \Theta _k}}}} \right)\mathcal{G}\left( {{\boldsymbol \Theta _k}} \right) + {\mathcal{F}}\left( {{\boldsymbol \Theta _k}} \right)} \right],
\end{equation}
\begin{equation}\label{childdis}
p\left( {{{\cal C}_{{\boldsymbol \Theta _k}}}|{\boldsymbol \Theta _k},{\boldsymbol \Theta _m}} \right) \propto \exp \left[ {{\mathcal{J}_{{\boldsymbol \Theta _k}}}\left( {{\boldsymbol \Theta _m},{{\cal C}_{{\boldsymbol \Theta _k}}}} \right)\mathcal{G}\left( {{\boldsymbol \Theta _k}} \right) + \mathcal{B}\left( {{{\cal C}_{{\boldsymbol \Theta _k}}}} \right)} \right],
\end{equation}
where $\mathcal{F}\left( \cdot \right)$ and $\mathcal{B}\left( \cdot \right)$ mean the functions only associated with the corresponding variables. ${{{\cal J}_{{\boldsymbol \Theta _k}}}\left( {{{\cal P}_{{\Theta _k}}}} \right)}$ and ${{{\cal J}_{{\boldsymbol \Theta _k}}}\left( {{\boldsymbol\Theta _m},{{\cal C}_{{\boldsymbol \Theta _k}}}} \right)}$ are the natural sufficient statistics of the corresponding exponential family distributions. Given the naturally sufficient statistics, the first moment and second moment of the distribution are known. The natural sufficient statistics expressions change with different variables and will be illustrated in the following subsection.

Substituting \eqref{parentdis} and \eqref{childdis} into \eqref{variationalphik}, it yields
\begin{equation}\label{eqvaria}
{q^{(\xi )}}\left( {{{\boldsymbol{\Theta }}_k}} \right) \propto \exp \left\{ {{{\cal Q}_{{{\boldsymbol{\Theta }}_k}}^{\left( {\xi} \right)}}\mathcal{G}\left( {{\boldsymbol \Theta _k}} \right) + \mathcal{F}\left( {{\boldsymbol \Theta _k}} \right)} \right\},
\end{equation}
where ${{\cal Q}_{{{\boldsymbol{\Theta}}_k}}^{\left( {\xi } \right)}}$ is the $\xi$-th estimated natural statistics of the distribution ${q^{(\xi )}}\left( {{{\boldsymbol{\Theta }}_k}} \right)$. Intuitively,  the iterative update of the variational distribution ${q^{(\xi )}}\left( {{{\boldsymbol{\Theta }}_k}} \right)$ only depends on the update of the term ${{\cal Q}_{{{\boldsymbol{\Theta}}_k}}^{\left( {\xi } \right)}}$ and it can be partitioned into two different terms:
\begin{equation}\label{variationalnatrualparamter}
\begin{aligned}
{\mathcal{Q}^{(\xi)}_{{{\boldsymbol{\Theta }}_k}}}& = \underbrace {{\mathbb{E}_{\prod\limits_{j \ne k} {q^{(\xi)}\left( {{{\boldsymbol{\Theta }}_j}} \right)} }}\left[ {{\mathcal{J}_{{\boldsymbol \Theta _k}}}\left( {\mathcal{P}_{{\boldsymbol \Theta}_k}} \right)} \right]}_{\mathcal{Q}^{\left(\xi \right)}_{{\mathcal{P}_{{\boldsymbol \Theta}_k}}\rightarrow{{\boldsymbol{\Theta }}_k}}} \\&+ \sum\limits_{\mathcal C_{{\boldsymbol{\Theta }}_k}} \underbrace {{{\mathbb{E}_{\prod\limits_{j \ne k} {q^{(\xi)}\left( {{{\boldsymbol{\Theta }}_j}} \right)} }}\left[ {{\mathcal{J}_{{\boldsymbol \Theta _k}}}\left( {{{\boldsymbol{\Theta }}_m},{\mathcal{C}_{{\boldsymbol \Theta}_k}}} \right)} \right]}}_{\mathcal{Q}^{\left(\xi \right)}_{{\mathcal{C}_{{\boldsymbol \Theta}_k}}\rightarrow{{\boldsymbol{\Theta }}_k}}},
\end{aligned}
\end{equation}
where the first subterm ${\mathcal{Q}^{\left(\xi \right)}_{{\mathcal{P}_{{\boldsymbol \Theta}_k}}\rightarrow{{\boldsymbol{\Theta }}_k}}}$ is the message from the parent node ${\mathcal{P}_{{\boldsymbol \Theta}_k}}$ and the second subterm ${\mathcal{Q}^{\left(\xi \right)}_{{\mathcal{C}_{{\boldsymbol \Theta}_k}}\rightarrow{{\boldsymbol{\Theta }}_k}}}$ is the message from the child node ${\mathcal{C}_{{\boldsymbol \Theta}_k}}$.

\begin{figure}
  \centering
  \includegraphics[width=2.5in]{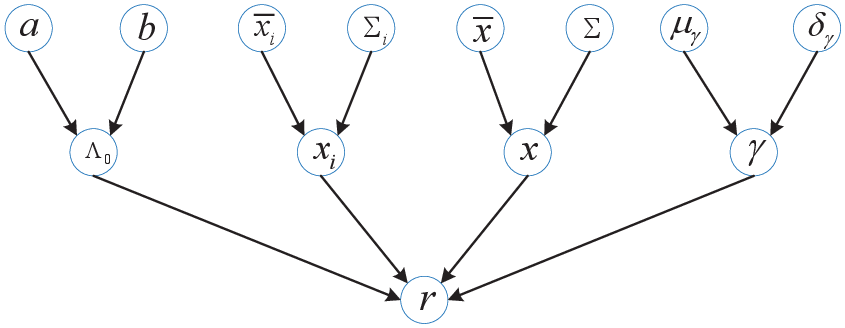}\\
  \caption{The probabilistic graph of the joint probability}\label{fig:graphics}
\end{figure}

Given the message update in \eqref{variationalnatrualparamter}, the original cumbersome distribution iterations are reduced to the lightweight message update between the nodes in the directed graph in Fig.\ref{fig:graphics}.  Specifically, the messages are essentially the expectations of the natural sufficient statistics of the prior distributions and the likelihood distribution.  Given the messages from the neighboring nodes, the sufficient statistics of the posterior distributions can be
updated iteratively. Given the message update scheme, the natural sufficient statistics of the variational distributions $q\left( {{\boldsymbol \Theta}_k}\right)$ are obtained iteratively. However,  it is too complicated to obtain closed-form solutions and the derivations of the messages are still non-trivial and still challenging due to the unknown fading parameter and the unknown distance-dependent noise. In the following subsections, the detailed theoretical derivations of the messages are presented.

\section{Joint Localization and Parameter Estimation Algorithm}

Using the message passing scheme, the unknown parameters are estimated via the updated messages from the parent nodes and the child nodes. In the following, we present the message update and the estimation of unknown parameters in detail.
\subsection{Estimation of path loss exponent parameter $\gamma$ }\label{estiofgamma}
To find the estimation of the unknown parameter $\gamma$,  it is required to get the estimation of the natural sufficient statistics $\mathcal{Q}_\gamma ^{\left( {\xi } \right)}$ of the variational distribution ${q^{\left( {\xi  } \right)}}\left( \gamma  \right)$. According to \eqref{variationalnatrualparamter}, the message $\mathcal{Q}_{{\mathcal{P}_\gamma } \to \gamma }^{\left( {\xi } \right)}$ from its parent nodes ${{\mu _\gamma }}$ and ${{\delta _\gamma }}$ and the message $\mathcal{Q}_{\boldsymbol r \to \gamma }^{\left( {\xi } \right)}$ from its child node $\boldsymbol r$ are derived respectively.

\textbf{Message $\mathcal{Q}_{{\mathcal{P}_\gamma } \to \gamma }^{\left( {\xi } \right)}$}: In Fig.\ref{fig:graphics},  the message $\mathcal{Q}_{{\mathcal{P}_\gamma } \to \gamma }^{\left( {\xi } \right)}$ from parent nodes ${{\mu _\gamma }}$ and ${{\delta _\gamma }}$ is related to the prior distribution ${\cal N}\left( {\gamma |{\mu _\gamma },{\delta _\gamma }} \right)$. By applying \eqref{parentdis} to the prior distribution of $\gamma$, the prior distribution of $\gamma$ can be rewritten as
 \begin{equation}\label{messQgamma}
p\left( \gamma  \right) = \exp \left( {\left[ {\begin{array}{*{20}{c}}
{\frac{{{\mu _\gamma }}}{{{\delta _\gamma }}}}&{ - \frac{1}{{2{\delta _\gamma }}}}
\end{array}} \right]\left[ {\begin{array}{*{20}{c}}
\gamma \\
{{\gamma ^2}}
\end{array}} \right]} \right) + {\cal C},
\end{equation}
Comparing with the expression in \eqref{parentdis}, we can obtain the natural sufficient statistics of $\gamma$ as ${\mathcal{J}_\mathcal{\gamma}}\left( {{\mathcal{P}_\gamma }} \right) = \left[ {\begin{array}{*{20}{c}}{\frac{{{\mu _\gamma }}}{{{\delta _\gamma }}}}&-{\frac{1}{{2{\delta _\gamma }}}}\end{array}} \right]$ and $\mathcal{G}\left( \gamma  \right) = \left[ {\begin{array}{*{20}{c}}\gamma \\{{\gamma ^2}}\end{array}} \right]$. By substituting the prior distribution of $\gamma$ into \eqref{variationalnatrualparamter} , the message  $\mathcal{Q}_{{\mathcal{P}_\gamma } \to \gamma }^{\left( {\xi } \right)}$ can be given by
\begin{equation}\label{mesagepgammatogamma}
\mathcal{Q}_{{{\cal P}_\gamma }{ \to  \gamma }}^{\left( \xi \right)} = \left[ {\begin{array}{*{20}{c}}
{\frac{{{\mu _\gamma }}}{{{\delta _\gamma }}}}&-{\frac{1}{{2{\delta _\gamma }}}}
\end{array}} \right].
\end{equation}

\textbf{Message $\mathcal{Q}_{\boldsymbol r \to \gamma }^{\left( {\xi } \right)}$}:  The message $\mathcal{Q}_{\boldsymbol r \to \gamma }^{\left( {\xi} \right)}$ from the child node $\boldsymbol r$ to node $\gamma$ involves the likelihood function in \eqref{likelihood} and the likelihood function with respect to the parameter $\gamma$ can be given by
    \begin{equation}\label{likelihoodpropo}
p\left( {{\boldsymbol{r}}\mid {\boldsymbol{\Theta }}} \right) \propto \exp \left( { -\sum _{i=1}^N \frac{\gamma }{2}\ln {d_i} - \sum _{i=1}^N\frac{{{{\left( {{r_i} - {d_i}} \right)}^2}{\Lambda _0}}}{{2d_i^\gamma }}} \right).
    \end{equation}

Due to the nonlinear unknown parameter $\gamma$ in the second term in \eqref{likelihoodpropo}, it is intractable to present the direct reformulation by following \eqref{childdis}. Hence, we apply Taylor expansion to linearize the likelihood function with respect to the parameter $\gamma$. By denoting $g\left( \gamma  \right) = \sum\limits_{i = 1}^N {{{\left( {{g_i}\left( \gamma  \right)} \right)}^2}} = \sum\limits_{i = 1}^N {{{\left( {\frac{{{r_i} - {d_i}}}{{d_i^{\frac{\gamma }{2}}}}} \right)}^2}} $, the Taylor expansion of $g\left( \gamma  \right)$ can be given by
\begin{equation}\label{Gammataylorexpan}
\begin{aligned}
&g\left( \gamma  \right) \approx \sum\limits_{i = 1}^N {{{\left( {{g_i}\left( {{\gamma ^{\left( \xi  \right)}}} \right) + {\boldsymbol{\Delta }}_{\gamma ,i}\left( {\gamma  - {\gamma ^{\left( \xi  \right)}}} \right)} \right)}^2}} \\
 &= \sum\limits_{i = 1}^N {{{\left( {{\boldsymbol{\Delta }}_{\gamma ,i}} \right)}^2}} {\gamma ^2} - 2\sum\limits_{i = 1}^N {{\boldsymbol{\Delta }}_{\gamma ,i}\left( {{\gamma ^{\left( \xi  \right)}} - {g_i}\left( {{\gamma ^{\left( \xi  \right)}}} \right) } \right)} \gamma  + \mathcal{C}\\
\end{aligned},
\end{equation}
where ${\boldsymbol{\Delta }}_{\gamma ,i}^{\left( \xi  \right)} =  - \frac{1}{2}\left( {{r_i} - {d_i}} \right)d_i^{ - \frac{\gamma }{2}}\ln {d_i}$ is the derivative with respect to $\gamma$.  Hence, substituting \eqref{Gammataylorexpan} into \eqref{childdis} and \eqref{likelihoodpropo}, it yields
  \begin{equation}\label{likelihoodpropo1}
    p\left( {{\boldsymbol{r}}\mid {\boldsymbol{\Theta }}} \right)\propto \exp \left(\left[ {\begin{array}{*{20}{c}}
{{\boldsymbol {\mathcal U}_\gamma }}&{-\frac{{1}}{2}}{\boldsymbol {\cal A} _\gamma }
\end{array}} \right]\left[ {\begin{array}{*{20}{c}}
\gamma \\
{{\gamma ^2}}
\end{array}} \right]\right),
  \end{equation}
where ${{\boldsymbol{\cal A}}_\gamma }  = \sum\limits_{i = 1}^N \frac{1}{4}{{{\left( {\frac{{{r_i} - {d_i}}}{{d_i^{{\gamma  \mathord{\left/
 { {2}} \right.
 }}}}}\ln {d_i}} \right)}^2}}\Lambda_0$ and ${{\boldsymbol{\cal U}}_\gamma } = \sum\limits_{i = 1}^N {{{\left( {{\boldsymbol{\Delta }}_{\gamma ,i}} \right)}^2}{{\boldsymbol{\Theta }}_{\gamma ,i}}}$ with ${\boldsymbol{\Theta }}_{\gamma ,i} = {{\gamma ^{\left( \xi  \right)}}{\Lambda _0} - {g_i}\left( {{\gamma ^{\left( \xi  \right)}}} \right){\Lambda _0} - \frac{1}{4}\ln {d_i}}$.

Therefore, the natural sufficient statistics of $\gamma$ in \eqref{likelihood} can be given by ${{\cal J}_{\gamma}}\left( \boldsymbol r \right) = \left[ {\begin{array}{*{20}{c}}{{\boldsymbol{\cal U}_\gamma }}&-{\frac{1}{2}{\boldsymbol{\mathcal A}_\gamma }}\end{array}} \right]$.  Then plugging the natural sufficient statistics into  \eqref{variationalnatrualparamter}, yields
\begin{equation}\label{mesagertogamma}
{\cal Q}_{{\boldsymbol{r}} \to \gamma }^{\left( {\xi } \right)} = {\mathbb{E}_{{q^{\left( \xi  \right)}}\left( {{\boldsymbol \Theta _{\backslash \gamma }}} \right)}}\left[ {{{\cal J}_{\gamma}}\left( \boldsymbol r \right)} \right] = \left[ {\begin{array}{*{20}{c}}
{\boldsymbol{\cal U}_\gamma ^{\left( \xi  \right)}}&-{\frac{1}{2}\boldsymbol{\mathcal A}_\gamma ^{\left( \xi  \right)}}
\end{array}} \right],
\end{equation}
where
\begin{equation}\label{Aiter}
\begin{aligned}
\boldsymbol {\cal A}_\gamma ^{\left( \xi  \right)} &= {\mathbb{E}_{{q^{\left( \xi  \right)}}\left( {{{\boldsymbol{\Theta }}_{\backslash \gamma }}} \right)}}\left( {{\boldsymbol {\cal A}_\gamma }} \right)  \approx \sum\limits_{i = 1}^N {{{\left( {\frac{{{r_i} - d_i^{\left( \xi  \right)}}}{{{{\left( {d_i^{\left( \xi  \right)}} \right)}^{\frac{{{\gamma ^{\left( \xi  \right)}}}}{2}}}}}\ln d_i^{\left( \xi  \right)}} \right)}^2}} \frac{{{a^{\left( \xi  \right)}}}}{{{b^{\left( \xi  \right)}}}}
\end{aligned}
\end{equation}
The approximation holds by applying the Taylor expansion at ${\boldsymbol x}^{\left( \xi  \right)}$ and ${\boldsymbol x}_i^{\left( \xi  \right)}$, which are obtained via the $\xi$-th estimation of target node location and sensor node locations respectively.
where ${a^{\left( \xi  \right)}}$ and ${b^{\left( \xi  \right)}}$ are the parameters in
${q^{\left( \xi  \right)}}\left( {{\Lambda _0}} \right) = \Gamma \left( {{\Lambda _0}|{a^{\left( \xi  \right)}},{b^{\left( \xi  \right)}}} \right)$ which will be introduced later in subsection III-B. The parameter $\boldsymbol {\cal U}_\gamma ^{\left( \xi  \right)}$ is given by
\begin{equation}\label{Uxi}
{\boldsymbol{\cal U}}_\gamma ^{\left( \xi  \right)} = {\mathbb{E}_{{q^{\left( \xi  \right)}}\left( {{{\boldsymbol{\Theta }}_{\backslash \gamma }}} \right)}}\left( {{{\boldsymbol{\cal U}}_\gamma }} \right) \approx \sum\limits_{i = 1}^N {{{\left( {{\boldsymbol{\Delta }}_{\gamma ,i}^{\left( \xi  \right)}} \right)}^2}{\boldsymbol{\Theta }}_{\gamma ,i}^{\left( \xi  \right)}},
\end{equation}
where ${\boldsymbol{\Theta }}_{\gamma ,i}^{\left( \xi  \right)} = {\gamma ^{\left( \xi  \right)}}\frac{{{a^{\left( \xi  \right)}}}}{{{b^{\left( \xi  \right)}}}} - {g_i}\left( {{\gamma ^{\left( \xi  \right)}}} \right)\frac{{{a^{\left( \xi  \right)}}}}{{{b^{\left( \xi  \right)}}}} - \frac{1}{4}\ln d_i^{\left( \xi  \right)}$ and ${\mathbb{E}_{{q^{\left( \xi  \right)}}\left( {{\boldsymbol \Theta _{\backslash \gamma }}} \right)}}\left( {\ln {d_i}} \right) = \ln d_i^{\left( \xi  \right)}$  in \cite{LiSecure20IoT} with $d_i^{\left( \xi  \right)} = {\left\| {{\boldsymbol x^{\left( \xi  \right)}} - \boldsymbol x_i^{\left( \xi  \right)}} \right\|_2}$.

Invoked by the conjugate prior principle, the variational distribution $q\left(\gamma\right)$ is also a Gamma distribution. By plugging \eqref{mesagepgammatogamma} and \eqref{Uxi} into \eqref{variationalnatrualparamter}, it yields
\begin{equation}\label{messagegamma}
\begin{aligned}
{\cal Q}_\gamma ^{(\xi )} &= \left[ {\begin{array}{*{20}{c}}
{\frac{{\mu _\gamma ^{\left( {\xi } \right)}}}{{\delta _\gamma ^{\left( {\xi } \right)}}}}&{ - \frac{1}{{2\delta _\gamma ^{\left( {\xi } \right)}}}}
\end{array}} \right] \\ &= \left[ {\begin{array}{*{20}{c}}
{\frac{{{\mu _\gamma }}}{{{\delta _\gamma }}} +\boldsymbol { \mathcal U}_\gamma ^{\left( \xi  \right)}}&{ - \frac{1}{{2{\delta _\gamma }}} - \frac{1}{2}\boldsymbol{\cal A}_\gamma ^{\left( \xi  \right)}}
\end{array}} \right].
\end{aligned}
\end{equation}

Hence, the $\xi$-th mean and variance of $\gamma$ can be respectively given by
\begin{equation}\label{gammavar}
\delta _\gamma ^{\left( {\xi } \right)} = {\left( {\frac{1}{{{\delta _\gamma }}} + \boldsymbol {\cal A}_\gamma ^{\left( \xi  \right)}} \right)^{ - 1}},
\end{equation}
\begin{equation}\label{gammamu}
\mu _\gamma ^{\left( {\xi } \right)} = \delta _\gamma ^{\left( {\xi } \right)}\left( {\frac{{{\mu _\gamma }}}{{{\delta _\gamma }}} + \boldsymbol {\mathcal U}_\gamma ^{\left( \xi  \right)}} \right).
\end{equation}

Thus, the $\xi$-th estimation of  $\gamma$ can be obtained via \eqref{gammamu}. Intuitively, the natural sufficient statistics $\mathcal{Q}_\gamma ^{\left( {\xi } \right)}$, $\mathcal{Q}_{{\mathcal{P}_\gamma } \to \gamma }^{\left( {\xi } \right)}$ and $\mathcal{Q}_{\boldsymbol r \to \gamma }^{\left( {\xi } \right)}$ are the means and variances of the their distributions and the messages are the expectations of means and variances. \vspace{-1em}
\subsection{Estimation of fading parameter $\Lambda_{0}$ }
In order to estimate the unknown parameter $\Lambda_{0}$,  it is required to estimate the natural sufficient statistics $\mathcal{Q}_{\Lambda_{0}}^{\left( {\xi  } \right)}$ of the variational distribution ${q^{\left( {\xi } \right)}}\left( {\Lambda_{0}}  \right)$.  Similarly, the message $\mathcal{Q}_{{\mathcal{P}_{\Lambda_{0}} } \to {\Lambda_{0}} }^{\left( {\xi } \right)}$ from its parent nodes ${{a  }}$ and ${{b}}$ and the message $\mathcal{Q}_{\boldsymbol r \to  {\Lambda_{0}} }^{\left( {\xi } \right)}$ from its child node $\boldsymbol r$ are also derived respectively.

\textbf{Message $\mathcal{Q}_{{\mathcal{P}_{\Lambda_{0}} } \to {\Lambda_{0}} }^{\left( {\xi } \right)}$}: The message $\mathcal{Q}_{{\mathcal{P}_{\Lambda_{0}} } \to {\Lambda_{0}} }^{\left( {\xi } \right)}$ involves the parent nodes $a$ and $b$ which is given in the prior distribution \eqref{gammadis}. By following \eqref{parentdis}, the prior distribution of ${\Lambda_{0}}$ can be rewritten as
      \begin{equation}\label{messQLambda}
\begin{aligned}
p\left( {\left. {{\Lambda _0}} \right|a,b} \right) &\propto \exp \left( {\left( {a - 1} \right)\ln {\Lambda _0} - b{\Lambda _0}} \right)\\
& = \exp \left( {\left[ {\begin{array}{*{20}{c}}
{a - 1}&-b
\end{array}} \right]\left[ {\begin{array}{*{20}{c}}
{{\Lambda _0}}\\
{\ln {\Lambda _0}}
\end{array}} \right]} \right),
\end{aligned}
      \end{equation}

Therefore,  the natural sufficient statistics of ${\Lambda_{0}}$ is given by ${\mathcal{J}_{\Lambda_{0}}}\left( {{\mathcal{P}_{\Lambda_{0}} }} \right) = {\left[ {\begin{array}{*{20}{c}}{a - 1}&b\end{array}} \right]}$ and $\mathcal{G}\left( {\Lambda_{0}}  \right) = \left[ {\begin{array}{*{20}{c}}{\Lambda_{0}} \\{{\ln {\Lambda_{0}}}}\end{array}} \right]$. By substituting the natural sufficient statistics of ${\Lambda_{0}}$ into \eqref{variationalnatrualparamter}, the message  $\mathcal{Q}_{{\mathcal{P}_{\Lambda_{0}} } \to {\Lambda_{0}} }^{\left( {\xi } \right)}$ can be given by
\begin{equation}\label{messageQLambda}
{\cal Q}_{{\mathcal{P}_{{\Lambda _0}}} \to {\Lambda _0}}^{\left( {\xi } \right)} = \left[ {\begin{array}{*{20}{c}}
{a - 1}&{ - b}
\end{array}} \right].
\end{equation}

\textbf{Message $\mathcal{Q}_{\boldsymbol r \to {\Lambda_{0}} }^{\left( {\xi } \right)}$}:  According to \eqref{childdis}, $\mathcal{Q}_{\boldsymbol r \to {\Lambda_{0}} }^{\left( {\xi }\right)}$ is the message from the child node $\boldsymbol r$ and requires the exponential factorization of the likelihood function in \eqref{likelihood}, which is given by
{\begin{equation}\label{messageQLambdalikelihood}
\begin{aligned}
p\left( {{\boldsymbol{r}}\mid {\boldsymbol{\Theta }}} \right) \propto &\exp \left( {\frac{N}{2} \ln {\Lambda _0}  - \sum\limits_{i = 1}^N {\frac{{{{\left( {{r_i} - {d_i}} \right)}^2}{\Lambda _0}}}{{2d_i^\gamma }}} } \right)\\
 &= \exp \left( {\left[ {\begin{array}{*{20}{c}}
{{\mathcal{I}_{{\Lambda _0}}}}&\frac{N}{2}
\end{array}} \right]\left[ {\begin{array}{*{20}{c}}
{{\Lambda _0}}\\
{\ln {\Lambda _0}}
\end{array}} \right]} \right) + {\cal C},
\end{aligned}
\end{equation}}
where ${{\cal J}_{\Lambda_{0}}}\left( \boldsymbol r \right) = {\left[ {\begin{array}{*{20}{c}}{{\mathcal{I}_{{\Lambda _0}}}}&\frac{N}{2}\end{array}} \right]}$ is the natural sufficient statistics of ${\Lambda_{0}}$ involved in the likelihood and ${\mathcal{I}_{{\Lambda _0}}} =  - \sum\limits_{i = 1}^N {\frac{{{{\left( {{r_i} - {d_i}} \right)}^2}}}{{2d_i^\gamma }}}$. According to \eqref{variationalnatrualparamter}, the $\mathcal{Q}_{\boldsymbol r \to {\Lambda_{0}} }^{\left( {\xi }\right)}$ is the variational expectation of the natural sufficient statistics of ${\Lambda_{0}}$. But the sufficient statistics contain the nonlinear term and we also apply the Taylor expansion at the $\xi$-th estimation of $\boldsymbol x$, ${\boldsymbol x}_i$ and $\gamma$. Thus, the message $\mathcal{Q}_{\boldsymbol r \to {\Lambda_{0}} }^{\left( {\xi } \right)}$ is given by
\begin{equation}\label{ExmessageQ}
\begin{aligned}
\mathcal{Q}_{\boldsymbol r \to {\Lambda _0}}^{\left( {\xi } \right)} = {\mathbb{E}_{{q^{\left( \xi  \right)}}\left( {{\boldsymbol \Theta _{\backslash {\Lambda _0}}}} \right)}}\left[ {{{\cal J}_{\Lambda _0}}\left( \boldsymbol r \right)} \right] = \left[ {\begin{array}{*{20}{c}}{\mathcal{I}_{{\Lambda _0}}^{\left( {\xi  + 1} \right)}}&\frac{N}{2}\end{array}} \right],
\end{aligned}
\end{equation}
where $\mathcal{I}_{{\Lambda _0}}^{\left( {\xi } \right)} =  - \sum\limits_{i = 1}^N {\frac{{{{\left( {{r_i} - d_i^{\left( \xi  \right)}} \right)}^2}}}{{2{{\left( {d_i^{\left( \xi  \right)}} \right)}^{{\gamma ^{\left( \xi  \right)}}}}}}}$ by using similar operations in \eqref{Gammataylorexpan}.

Substituting \eqref{ExmessageQ} and \eqref{messQLambda} into \eqref{variationalnatrualparamter} yields,
\begin{equation}\label{LambdaQ}
\begin{aligned}
{\cal Q}_{{\Lambda _0}}^{(\xi )} = \mathcal{Q}_{{{\cal P}_{{\Lambda _0}}} \to {\Lambda _0}}^{\left( {\xi } \right)} + {\cal Q}_{{\boldsymbol{r}} \to {\Lambda _0}}^{\left( {\xi } \right)}
 = \left[ {\begin{array}{*{20}{c}}
{a - 1 +\mathcal{ I}_{{\Lambda _0}}^{\left( {\xi } \right)}}{ - b + \frac{N}{2}}-1
\end{array}} \right].
\end{aligned}
\end{equation}

Thus,  the $\xi $-th variational distribution of $\Lambda_0$ can be given by ${q^{\left( {\xi } \right)}}\left( {{\Lambda _0}|{a^{\left( {\xi } \right)}},{b^{\left( {\xi } \right)}}} \right)$. Combining it with  \eqref{LambdaQ}, we can obtain ${a^{\left( {\xi  } \right)}} = a + \mathcal{I}_{{\Lambda _0}}^{\left( {\xi } \right)}$ and ${b^{\left( {\xi } \right)}} = b + \frac{N}{2}$ respectively.

Therefore, the $\xi$-th estimation of $\Lambda_0$ is given by
\begin{equation}\label{LambdaEstimation}
\Lambda _0^{\left( {\xi} \right)} = \frac{{{a^{\left( {\xi } \right)}}}}{{{b^{\left( {\xi } \right)}}}}.
\end{equation}

Different from the messages consisting of means and variances in subsection \ref{estiofgamma},  the messages of $\Lambda_0$ are comprised of expectations of statistics parameters from the prior distribution in \eqref{gammadis} and the likelihood distribution in \eqref{likelihood}.  These messages update the statistics parameters leading to the iterative estimation of  $\Lambda_0$.  

\subsection{Estimation of target node location ${\boldsymbol x}$ }
In this subsection, we focus on the estimation of the target node location ${\boldsymbol x}$. According to \eqref{variationalnatrualparamter}, the target node location estimation requires to iteratively update the natural sufficient statistics $\mathcal{Q}_{{\boldsymbol x}}^{\left( {\xi } \right)}$ of the variational distribution ${q^{\left( {\xi } \right)}}\left( {\boldsymbol x} \right)$.  Meanwhile, the message $\mathcal{Q}_{{\mathcal{P}_{{\boldsymbol x}} } \to {\boldsymbol x} }^{\left( {\xi } \right)}$ from its parent nodes $ \bar {\boldsymbol x}$ and ${\boldsymbol \Sigma}$ and the message $\mathcal{Q}_{\boldsymbol r \to  {\Lambda_{0}} }^{\left( {\xi } \right)}$ from its child node $\boldsymbol r$ are also derived respectively.

\textbf{Message $\mathcal{Q}_{{\mathcal{P}_{{\boldsymbol x}} } \to {\boldsymbol x} }^{\left( {\xi } \right)}$}:  According to the directed graph in Fig.\ref{fig:graphics}, the $\mathcal{Q}_{{\mathcal{P}_{{\boldsymbol x}} } \to {\boldsymbol x} }^{\left( {\xi  } \right)}$ is from its parent nodes $ \bar {\boldsymbol x}$ and ${\boldsymbol \Sigma}$, which is generated from the prior distribution of ${\boldsymbol x}$. Substituting the prior distribution of ${\boldsymbol x}$ into \eqref{parentdis} and simple trace manipulations, we can obtain
      \begin{equation}\label{priorX}
      \begin{aligned}
p\left( {\boldsymbol x} \right) \propto &\exp \left( {tr\left( { - \frac{1}{2}{{\left( {{\boldsymbol x} - \bar {\boldsymbol x}} \right)}^T}{\boldsymbol \Sigma ^{ - 1}}\left( {{\boldsymbol x} - \bar {\boldsymbol x}} \right)} \right)} \right)\\
 \propto& \exp \left( {tr\left( {\left[ {\begin{array}{*{20}{c}}
{{{\left( {\bar {\boldsymbol x}} \right)}^T}\boldsymbol \Sigma }&{ - \frac{1}{2}\boldsymbol \Sigma^{ - 1}}
\end{array}} \right]\left[ {\begin{array}{*{20}{c}}
{\boldsymbol x}\\
{{\boldsymbol x}{{\boldsymbol x}^T}}
\end{array}} \right]} \right)} \right),
\end{aligned}
      \end{equation}
where $tr$ means the trace of a matrix and natural sufficient statistics of ${\boldsymbol x}$ in the prior distribution is given by ${{\cal J}_{{\boldsymbol x}}}\left( {{{\cal P}_{\boldsymbol x}}} \right) = \left[ {\begin{array}{*{20}{c}}{{{\left( {\bar {\boldsymbol x}} \right)}^T}\boldsymbol \Sigma ^{ - 1}}&{ - \frac{1}{2}\boldsymbol \Sigma ^{ - 1}}\end{array}} \right]$ with ${\cal G}\left( {\boldsymbol x} \right) = \left[ {\begin{array}{*{20}{c}}{\boldsymbol x}\\{{\boldsymbol x}{{\boldsymbol x}^T}}\end{array}} \right]$. By substituting the natural sufficient statistics of ${\boldsymbol x}$ into \eqref{variationalnatrualparamter}, the message  $\mathcal{Q}_{{\mathcal{P}_{{\boldsymbol x}} } \to {\boldsymbol x} }^{\left( {\xi +1} \right)}$ can be given by
 \begin{equation}\label{messageQx}
{\cal Q}_{{{\cal P}_{\boldsymbol{x}}} \to {\boldsymbol{x}}}^{\left( {\xi} \right)} = \left[ {\begin{array}{*{20}{c}}
{{{\left( {\bar {\boldsymbol x}} \right)}^T}\boldsymbol \Sigma ^{ - 1}}&{ - \frac{1}{2}\boldsymbol \Sigma ^{ - 1}}
\end{array}} \right].
 \end{equation}

 \textbf{Message $\mathcal{Q}_{\boldsymbol r \to {\boldsymbol x} }^{\left( {\xi} \right)}$}:  In \eqref{childdis}, the message $\mathcal{Q}_{\boldsymbol r \to {\boldsymbol x} }^{\left( {\xi } \right)}$  is derive from the likelihood function and the natural sufficient statistics requires linear combination of the involved unknown parameter in the likelihood function.  Thus, we linearize the likelihood function with respect to the target node location ${\boldsymbol x}$. Firstly, the likelihood function can be rewritten as
    \begin{equation}\label{likeQ}
\ln p\left( {{\boldsymbol{r}}\mid {\boldsymbol{\Theta }}} \right) \propto  - \sum\limits_{i = 1}^N {\frac{\gamma }{2}} \ln {d_i} - \sum\limits_{i = 1}^N {\frac{{{{\left( {{r_i} - {d_i}} \right)}^2}{\Lambda _0}}}{{2d_i^\gamma }}},
    \end{equation}
  {where the likelihood function involves two nonlinear terms.} By denoting ${h_1}\left( {{\boldsymbol{x}},{{\boldsymbol{x}}_i}} \right) =  \gamma  \ln {d_i}$,  ${h_2}\left( {{\boldsymbol{x}},{{\boldsymbol{x}}_i}} \right) = {\frac{{{r_i} - {d_i}}}{{d_i^{\gamma /2}}}}$ and applying Taylor expansion at $\left( {{{\boldsymbol{x}}^{\left( {\xi } \right)}},{\boldsymbol{x}}_i^{\left( {\xi } \right)}} \right)$ which are respectively given by
  \begin{equation}\label{h1}
  \begin{aligned}
{h_1}\left( {{\boldsymbol{x}},{{\boldsymbol{x}}_i}} \right) ={{\gamma ^{\left( \xi  \right)}}} \ln d_i^{^{\left( \xi  \right)}} + {\left( {{{\boldsymbol{\omega }}^{\left( \xi  \right)}_i}} \right)^T}\left( {{\boldsymbol{x}} - {{\boldsymbol{x}}^{\left( \xi  \right)}}} \right),
  \end{aligned}
  \end{equation}
  where ${{\boldsymbol{\omega }}^{\left( \xi  \right)}_i} = {{\gamma ^{\left( \xi  \right)}}\frac{1}{{d_i^{^{\left( \xi  \right)}}}}{{\left. {\frac{{\partial {d_i}}}{{\partial {\boldsymbol{x}}}}} \right|}_{{{\boldsymbol{x}}^{\left( \xi  \right)}},{\boldsymbol{x}}_i^{\left( \xi  \right)}}}}$
and
  \begin{equation}\label{h2}
{h_2}\left( {{\boldsymbol{x}},{{\boldsymbol{x}}_i}} \right) = {\Upsilon ^{\left( \xi  \right)}_i} + {\left( {{\boldsymbol{\Gamma }}_i^{\left( \xi  \right)}} \right)^T}\left( {{\boldsymbol{x}} - {{\boldsymbol{x}}^{\left( \xi  \right)}}} \right),
  \end{equation}
where ${\Upsilon ^{\left( \xi  \right)}_i}$ and ${\boldsymbol{\Gamma }}_i^{\left( \xi  \right)} $ are respectively given by
\begin{equation}\label{h21}
{\Upsilon ^{\left( \xi  \right)}_i} =  {\frac{{{r_i} - d_i^{\left( \xi  \right)}}}{{{{\left( {d_i^{\left( \xi  \right)}} \right)}^{{\gamma ^{\left( \xi  \right)}}/2}}}}},
\end{equation}
\begin{equation}\label{h22}
{\boldsymbol{\Gamma }}_i^{\left( \xi  \right)} =  - \frac{{1 + 0.5\gamma d_i^{ - 1}\left( {{r_i} - {d_i}} \right)}}{{d_i^{\gamma /2}}}{\left. {\frac{{\partial {d_i}}}{{\partial {\boldsymbol{x}}}}} \right|_{{{\boldsymbol{x}}^{\left( \xi  \right)}},{\boldsymbol{x}}_i^{\left( \xi  \right)}}}.
\end{equation}

Substituting \eqref{h1} and \eqref{h2} into \eqref{likeQ} and taking exponential operation with both sides, it yields
\begin{equation}\label{likeQ1}
\begin{aligned}
p\left( {{\boldsymbol{r}}\mid {\boldsymbol{\Theta }}} \right) \propto \exp \left( {tr\left( { - \frac{1}{2}{{\boldsymbol{x}}^T}{\boldsymbol{\Delta }}_{\boldsymbol{x}}^{\left( \xi  \right)}{\boldsymbol{x}} + { {{\boldsymbol U}_{\boldsymbol{x}}^{\left( \xi  \right)}}}{\boldsymbol{x}}} \right)} \right)\\
 = \exp \left( {tr\left( {\left[ {\begin{array}{*{20}{c}}
{{{{\boldsymbol U}_{\boldsymbol{x}}^{\left( \xi  \right)}}}}&{ - \frac{1}{2}{\boldsymbol{\Delta }}_{\boldsymbol{x}}^{\left( \xi  \right)}}
\end{array}} \right]\left[ {\begin{array}{*{20}{c}}
{\boldsymbol{x}}\\
{{\boldsymbol{x}}{{\boldsymbol{x}}^T}}
\end{array}} \right]} \right)} \right)
\end{aligned}
\end{equation}
where ${{\cal J}_{\boldsymbol{x}}}\left( {\boldsymbol{r}} \right) = \left[ {\begin{array}{*{20}{c}}
{{\boldsymbol U}_{\boldsymbol{x}}^{\left( \xi  \right)}}&{ - \frac{1}{2}{\boldsymbol{\Delta }}_{\boldsymbol{x}}^{\left( \xi  \right)}}
\end{array}} \right]$ is the natural sufficient statistics of ${\boldsymbol x}$ involved in the likelihood function and ${\boldsymbol{\Delta }}_{\boldsymbol{x}}^{\left( \xi  \right)}$ and  ${{\boldsymbol U}_{\boldsymbol{x}}^{\left( \xi  \right)}}$ are respectively given by
\begin{equation}\label{Delta}
{\boldsymbol{\Delta }}_{\boldsymbol{x}}^{\left( \xi  \right)} = \Lambda_0 \sum\limits_{i = 1}^N {{{\left( {{\boldsymbol{\Gamma }}_i^{\left( \xi  \right)}} \right)}^T}{\boldsymbol{\Gamma }}_i^{\left( \xi  \right)}},
\end{equation}
\begin{equation}\label{U}
\begin{aligned}
{\boldsymbol U}_{\boldsymbol{x}}^{\left( \xi  \right)}& = \Lambda_0 {\left( {{{\boldsymbol{x}}^{\left( \xi  \right)}}} \right)^T}\sum\limits_{i = 1}^N {{{\left( {{\boldsymbol{\Gamma }}_i^{\left( \xi  \right)}} \right)}^T}{\boldsymbol{\Gamma }}_i^{\left( \xi  \right)}} - \sum\limits_{i = 1}^N  \Lambda_0{\left( {{\boldsymbol{\Gamma }}_i^{\left( \xi  \right)}} \right)^T}{\Upsilon ^{\left( \xi  \right)}_i} \\& + \sum\limits_{i = 1}^N  \Lambda_0{\left( {{{\boldsymbol{\omega }}^{\left( \xi  \right)}_i}} \right)^T}.
\end{aligned}
\end{equation}

The message $\mathcal{Q}_{\boldsymbol r \to {\boldsymbol x} }^{\left( {\xi } \right)}$ can be obtained as
\begin{equation}\label{messageQrx}
{\cal Q}_{{\boldsymbol{r}} \to {\boldsymbol{x}}}^{\left( {\xi} \right)} = {\mathbb{E}_{{q^{\left( \xi  \right)}}\left( {{{\boldsymbol{\Theta }}_{\backslash {{\boldsymbol x}}}}} \right)}}\left[ {{\mathcal{J}_{\boldsymbol{x}}}\left( {\boldsymbol{r}} \right)} \right] = \left[ {\begin{array}{*{20}{c}}
{\mathcal{U}_{\boldsymbol{x}}^{\left( {\xi} \right)}}&{ - \frac{1}{2}\mathcal{A}_{\boldsymbol{x}}^{\left( {\xi} \right)}}
\end{array}} \right],
\end{equation}
where ${\mathcal{U}_{\boldsymbol{x}}^{\left( {\xi } \right)}}$ and $\mathcal{A}_{\boldsymbol{x}}^{\left( {\xi } \right)}$ are  the variational expectations and obtained by replacing $\Lambda_0$ in ${\boldsymbol U}_{\boldsymbol{x}}^{\left( \xi  \right)}$ and ${\boldsymbol{\Delta }}_{\boldsymbol{x}}^{\left( \xi  \right)}$ with the $\xi$-th estimation $\Lambda^{\left(\xi\right)}_0$.

Plugging the messages ${\cal Q}_{{\boldsymbol{r}} \to {\boldsymbol{x}}}^{\left( {\xi} \right)}$ and $\mathcal{Q}_{{\mathcal{P}_{{\boldsymbol x}} } \to {\boldsymbol x} }^{\left( {\xi } \right)}$ into \eqref{variationalnatrualparamter}, it yields
\begin{equation}\label{nasofx}
\begin{aligned}
{\cal Q}_{\boldsymbol{x}}^{\left( \xi  \right)} &= {\cal Q}_{{\boldsymbol{r}} \to {\boldsymbol{x}}}^{\left( \xi  \right)} + {\cal Q}_{{{\cal P}_{\boldsymbol{x}}} \to {\boldsymbol{x}}}^{\left( \xi  \right)} \\&= \left[ {\begin{array}{*{20}{c}}
{{{\left( {{{\boldsymbol{x}}^{\left( \xi  \right)}}} \right)}^T}{{\left( {{{\boldsymbol{\Sigma }}^{\left( \xi  \right)}}} \right)}^{ - 1}}}&{ - \frac{1}{2}{{\left( {{{\boldsymbol{\Sigma }}^{\left( \xi  \right)}}} \right)}^{ - 1}}}
\end{array}} \right].
\end{aligned}
\end{equation}
where ${{{\boldsymbol{x}}^{\left( \xi  \right)}}}$ and ${{{\boldsymbol{\Sigma }}^{\left( \xi  \right)}}}$ are the $\xi$-th mean vector and covariance matrix respectively and are given by
\begin{equation}\label{xithofxmean}
{{\boldsymbol{x}}^{\left( \xi  \right)}} = \left( {\mathcal{U}_{\boldsymbol{x}}^{\left( \xi  \right)} + {{\left( {\overline {\boldsymbol{x}} } \right)}^T}{{\boldsymbol{\Sigma }}^{ - 1}}} \right){{\boldsymbol{\Sigma }}^{\left( \xi  \right)}},
\end{equation}
\begin{equation}\label{xithofxSigma}
{{\boldsymbol{\Sigma }}^{\left( \xi  \right)}} = {\left( {\mathcal{A}_{\boldsymbol{x}}^{\left( \xi  \right)} + {{\boldsymbol{\Sigma }}^{ - 1}}} \right)^{ - 1}}.
\end{equation}

Thus, the $\xi$-th estimation of target node location can be given by \eqref{xithofxmean}.

\subsection{Estimation of sensor node location ${\boldsymbol x}_i$ }
In the paper, the sensor node locations are not perfectly known and can be refined by using the message-passing algorithm.  In \eqref{variationalnatrualparamter}, the natural sufficient statistics $\mathcal{Q}_{{\boldsymbol x}_i }^{\left( {\xi } \right)}$ of the variational distribution ${q^{\left( {\xi } \right)}}\left( {\boldsymbol x}_i \right)$ is required and calculated based on the messages from its child nodes and parent nodes.
Hence, the message $\mathcal{Q}_{{\mathcal{P}_{{\boldsymbol x}_i } } \to {\boldsymbol x}_i  }^{\left( {\xi } \right)}$ from its parent nodes $ \bar{\boldsymbol x}_i $ and ${\boldsymbol \Sigma}_i$ and the message $\mathcal{Q}_{\boldsymbol r \to  {\boldsymbol x}_i  }^{\left( {\xi } \right)}$ from its child node $\boldsymbol r$ are derived respectively.

\textbf{Message $\mathcal{Q}_{{\mathcal{P}_{{\boldsymbol x}_i} } \to {\boldsymbol x}_i }^{\left( {\xi } \right)}$}: The message $\mathcal{Q}_{{\mathcal{P}_{{\boldsymbol x}_i} } \to {\boldsymbol x}_i }^{\left( {\xi } \right)}$ from the parent nodes are based on the prior distribution in \eqref{sensoruncertain} and according to  \eqref{parentdis}, we can obtain
         \begin{equation}\label{priorXi}
      \begin{aligned}
& p\left( {\boldsymbol x}_i \right) \propto  \exp \left( {tr\left( {\left[ {\begin{array}{*{20}{c}}
{{{\left( {\bar {\boldsymbol x}_i } \right)}^T}\boldsymbol \Sigma_i  {\boldsymbol x}^{ - 1}_i }&{ - \frac{1}{2}\boldsymbol \Sigma^{ - 1}_i }
\end{array}} \right]\left[ {\begin{array}{*{20}{c}}
{\boldsymbol x}_i \\
{{\boldsymbol x}_i {{\boldsymbol x}^T_i }}
\end{array}} \right]} \right)} \right),
\end{aligned}
      \end{equation}
Similarly, by substituting the natural sufficient statistics of ${\boldsymbol x}_i $ into \eqref{variationalnatrualparamter}, the message  $\mathcal{Q}_{{\mathcal{P}_{{\boldsymbol x}_i} } \to {\boldsymbol x}_i }^{\left( {\xi } \right)}$ can be given by
 \begin{equation}\label{messageQxi}
{\cal Q}_{{{\cal P}_{\boldsymbol{x}_i}} \to {\boldsymbol{x}}_i}^{\left( {\xi} \right)} = \left[ {\begin{array}{*{20}{c}}
{{{\left( {\bar {\boldsymbol x}_i} \right)}^T}\boldsymbol \Sigma ^{ - 1}_i}&{ - \frac{1}{2}\boldsymbol \Sigma ^{ - 1}_i}
\end{array}} \right].
 \end{equation}

 \textbf{Message $\mathcal{Q}_{\boldsymbol r \to {\boldsymbol x}_i }^{\left( {\xi } \right)}$}:  The derivations of message $\mathcal{Q}_{\boldsymbol r \to {\boldsymbol x}_i }^{\left( {\xi } \right)}$  is similar to the ones of message $\mathcal{Q}_{\boldsymbol r \to {\boldsymbol x} }^{\left( {\xi} \right)}$.  Applying Taylor expansion on ${h_1}\left( {{\boldsymbol{x}},{{\boldsymbol{x}}_i}} \right)$ and ${h_2}\left( {{\boldsymbol{x}},{{\boldsymbol{x}}_i}} \right)$ at ${\boldsymbol{x}}_i^{\left( {\xi } \right)}$, it yields
  \begin{equation}\label{h1xi}
  \begin{aligned}
{h_1}\left( {{\boldsymbol{x}},{{\boldsymbol{x}}_i}} \right) ={{\gamma ^{\left( \xi  \right)}}} \ln d_i^{^{\left( \xi  \right)}} - {\left( {{{\boldsymbol{\omega }}^{\left( \xi  \right)}_i}} \right)^T}\left( {{\boldsymbol{x}}_i - {{\boldsymbol{x}}_i^{\left( \xi  \right)}}} \right),
  \end{aligned}
  \end{equation}
and
  \begin{equation}\label{h2xi}
{h_2}\left( {{\boldsymbol{x}},{{\boldsymbol{x}}_i}} \right) = {\Upsilon ^{\left( \xi  \right)}_i} - {\left( {{\boldsymbol{\Gamma }}_i^{\left( \xi  \right)}} \right)^T}\left( {{\boldsymbol{x}}_i - {{\boldsymbol{x}}_i^{\left( \xi  \right)}}} \right),
  \end{equation}

Substituting \eqref{h1xi} and \eqref{h2xi} into \eqref{likeQ}, it yields
\begin{equation}\label{likeQ3}
\begin{aligned}
p\left( {{\boldsymbol{r}}\mid {\boldsymbol{\Theta }}} \right) \propto \exp \left( {tr\left( { - \frac{1}{2}{\boldsymbol{x}}_i^T{\boldsymbol{\Delta }}_{{{\boldsymbol{x}}_i}}^{\left( \xi  \right)}{{\boldsymbol{x}}_i} + {{\boldsymbol U}}_{{{\boldsymbol{x}}_i}}^{\left( \xi  \right)}{{\boldsymbol{x}}_i}} \right)} \right)\\
 = \exp \left( {tr\left( {\left[ {\begin{array}{*{20}{c}}
{{{\boldsymbol U}}_{{{\boldsymbol{x}}_i}}^{\left( \xi  \right)}}&{ - \frac{1}{2}{\boldsymbol{\Delta }}_{{\boldsymbol{x}}_i}^{\left( \xi  \right)}}
\end{array}} \right]\left[ {\begin{array}{*{20}{c}}
{{{\boldsymbol{x}}_i}}\\
{{{\boldsymbol{x}}_i}{{\boldsymbol{x}}_i}^T}
\end{array}} \right]} \right)} \right),
\end{aligned}
\end{equation}
where ${{\cal J}_{{{\boldsymbol{x}}_i}}}\left( {\boldsymbol{r}} \right) = \left[ {\begin{array}{*{20}{c}}{{{{\boldsymbol U}}_{{{\boldsymbol{x}}_i}}^{\left( \xi  \right)}} }&{ - \frac{1}{2}{\boldsymbol{\Delta }}_{{\boldsymbol{x}}_i}^{\left( \xi  \right)}}
\end{array}} \right]$ with ${{{\boldsymbol U}}_{{{\boldsymbol{x}}_i}}^{\left( \xi  \right)}} = \Lambda_0 \left({\left( {{{\boldsymbol{x}}_i^{\left( \xi  \right)}}} \right)^T}{{{\left( {{\boldsymbol{\Gamma }}_i^{\left( \xi  \right)}} \right)}^T}{\boldsymbol{\Gamma }}_i^{\left( \xi  \right)}}+ {\left( {{\boldsymbol{\Gamma }}_i^{\left( \xi  \right)}} \right)^T}{\Upsilon ^{\left( \xi  \right)}_i} -  {\left( {{{\boldsymbol{\omega }}^{\left( \xi  \right)}_i}} \right)^T}\right)$ and ${\boldsymbol{\Delta }}_{{\boldsymbol{x}}_i}^{\left( \xi  \right)}= \Lambda_0 {{{\left( {{\boldsymbol{\Gamma }}_i^{\left( \xi  \right)}} \right)}^T}{\boldsymbol{\Gamma }}_i^{\left( \xi  \right)}}$.

In a analogous way to the message $\mathcal{Q}_{\boldsymbol r \to {\boldsymbol x} }^{\left( {\xi} \right)}$, we can obtain the message $\mathcal{Q}_{\boldsymbol r \to {\boldsymbol x}_i }^{\left( {\xi} \right)}$  as
\begin{equation}\label{messageQrxi}
\mathcal{Q}_{{\boldsymbol{r}} \to {{\boldsymbol{x}}_i}}^{\left( \xi  \right)} = {\mathbb{E}_{{q^{\left( \xi  \right)}}\left( {{{\boldsymbol{\Theta }}_{\backslash {{\boldsymbol{x}}_i}}}} \right)}}\left[ {{{\cal J}_{{{\boldsymbol{x}}_i}}}\left( {\boldsymbol{r}} \right)} \right] = \left[ {\begin{array}{*{20}{c}}
{{\cal U}_{{{\boldsymbol{x}}_i}}^{\left( \xi  \right)}}&{ - \frac{1}{2}{\cal A}_{{{\boldsymbol{x}}_i}}^{\left( \xi  \right)}}
\end{array}} \right],
\end{equation}

${\mathcal{U}_{\boldsymbol{x}_i}^{\left( {\xi } \right)}}$ and $\mathcal{A}_{\boldsymbol{x}_i}^{\left( {\xi } \right)}$ are also obtained by replacing $\Lambda_0$ with the $\xi$-th estimation $\Lambda^{\left(\xi\right)}_0$.

Plugging the messages ${\cal Q}_{{\boldsymbol{r}} \to {\boldsymbol{x}_i}}^{\left( {\xi} \right)}$ and $\mathcal{Q}_{{\mathcal{P}_{{\boldsymbol x}_i} } \to {\boldsymbol x}_i }^{\left( {\xi } \right)}$ into \eqref{variationalnatrualparamter}, it yields
\begin{equation}\label{nasofxi}
\begin{aligned}
{\cal Q}_{\boldsymbol{x}_i}^{\left( \xi  \right)} & = {\cal Q}_{{\boldsymbol{r}} \to {\boldsymbol{x}_i}}^{\left( \xi  \right)} + {\cal Q}_{{{\cal P}_{\boldsymbol{x}_i}} \to {\boldsymbol{x}_i}}^{\left( \xi  \right)}\\& = \left[ {\begin{array}{*{20}{c}}
{{{\left( {{{\boldsymbol{x}}^{\left( \xi  \right)}_i}} \right)}^T}{{\left( {{{\boldsymbol{\Sigma }}^{\left( \xi  \right)}_i}} \right)}^{ - 1}}}&{ - \frac{1}{2}{{\left( {{{\boldsymbol{\Sigma }}^{\left( \xi  \right)}_i}} \right)}^{ - 1}}}
\end{array}} \right].
\end{aligned}
\end{equation}
where ${{{\boldsymbol{x}}^{\left( \xi  \right)}_i}}$ and ${{{\boldsymbol{\Sigma }}^{\left( \xi  \right)}_i}}$ are the $\xi$-th mean vector and covariance matrix respectively and are given by
\begin{equation}\label{xthofximean}
{\boldsymbol{x}}_i^{\left( \xi  \right)} = \left( {\mathcal{U}_{{{\boldsymbol{x}}_i}}^{\left( \xi  \right)} + {{\left( {{{{\boldsymbol{\bar x}}}_i}} \right)}^T}{\boldsymbol{\Sigma }}_i^{ - 1}} \right){\boldsymbol{\Sigma }}_i^{\left( \xi  \right)},
\end{equation}
\begin{equation}\label{xthofxiSigma}
{\boldsymbol{\Sigma }}_i^{\left( \xi  \right)} = {\left( {\mathcal{A}_{{{\boldsymbol{x}}_i}}^{\left( \xi  \right)} + {\boldsymbol{\Sigma }}_i^{ - 1}} \right)^{ - 1}}.
\end{equation}

Thus, the $\xi$-th estimation of target node location can be given by \eqref{xthofximean}.

\section{Discussions}

The proposed joint estimation algorithm is meticulously crafted within the framework of variational Bayesian learning and makes use of a probabilistic graphical model. Operating as an iterative message-passing algorithm, the convergence of our proposed method hinges on the intricacies of the directed probabilistic graph \cite{BinTWCParametric22}. As depicted in Fig.\ref{fig:graphics}, the messages generated by our algorithm traverse through a free-loop graph, a structure that has been theoretically demonstrated to ensure convergence. This architectural design not only adheres to established theoretical principles in probabilistic graphical models but also underscores the reliability and effectiveness of our algorithm.

In \textbf{Algorithm I}, the computational intricacy primarily originates from the inversion of covariance matrices during the estimation of both the target node location and sensor node locations. Notably, the estimation of other unknown parameters involves scalar operators, contributing negligibly to the overall complexity. The specific complexity of inverting the covariance matrices in equations \eqref{xithofxSigma} and \eqref{xthofxiSigma} is characterized by ${{\mathcal{O}}}\left( {8\left( {M + 1} \right)} \right)$. Consequently, the cumulative computational complexity of \textbf{Algorithm I} is proportional to ${{\mathcal{O}}}\left( {8\left( {M + 1} \right)} \right)$. In comparison to existing methods, as delineated in \cite{WCLLiRobust21}, the computational complexities of  ML-GMP, and IRGLS adhere to proportional relationships of ${{\mathcal{O}}}\left( {M} \right)$, and ${\cal O}\left({\left( {\left( {\frac{{{M}}}{{{2}}}{{ + 1}}} \right)\left( {{{M - 1}}} \right)} \right)^{\rm{3}}}{\rm{ + 12}}{\left( {\left( {\frac{{{M}}}{{\rm{2}}}{\rm{ + 1}}} \right)\left( {{{M - 1}}} \right)} \right)^{\rm{2}}}\right)$, respectively. From an intuitive standpoint, \textbf{Algorithm I} exhibits complexities similar to ML-GMP, being on a similar scale, and surpasses the other algorithm in terms of efficiency.

{
\begin{algorithm}[htbp]
  \caption{Joint Localization and Parameter Estimation Algorithm}
  \begin{algorithmic}[1]
      \State Input the parameters $N$, ${\mathcal T}$, $a$, $b$, $\mu_{\gamma}$, $\delta_{\gamma}$, $\bar {\boldsymbol x}$, $\boldsymbol \Sigma$, $\bar {\boldsymbol x}_i$, ${\boldsymbol \Sigma}_i$, ${\boldsymbol p}_u$, ${\boldsymbol p}_r$ and the involved distributions $p\left( {\left. {{\Lambda _0}} \right|a,b} \right)$, $\mathcal{N}\left( {\gamma |{\mu _\gamma },{\delta _\gamma }} \right)$, $\mathcal{N}\left( {{\boldsymbol{x}}|{\boldsymbol{\bar x}},{\boldsymbol{\Sigma }}} \right)$ and $\mathcal{N}\left( {{\boldsymbol{x}}_i|{\boldsymbol{\bar x}}_i,{\boldsymbol{\Sigma }}_i} \right)$
       and collect measurements $\boldsymbol r$;
      \State $\xi = 1$ and calculate ${{\cal{D}}^{\left( 0 \right)}}$ with initial distributions;
      \While {$\left| {{{\cal{D}}^{\left( {\xi} \right)}} - {{\cal{D}}^{\left( \xi-1 \right)}}} \right| > {\cal T}$}
      \State $\xi = \xi + 1;$
      \State ${{\cal{D}}^{\left( \xi  \right)}}={\text{  KL}}\left( {{q^{\left( \xi  \right)}}\left( {\boldsymbol{\Theta }} \right)||p\left( {{\boldsymbol{\Theta }}|{\boldsymbol{r}}} \right)} \right)$;
      \State Updating the messages $\mathcal{Q}_{{\mathcal{P}_\gamma } \to \gamma }^{\left( {\xi } \right)}$ in \eqref{mesagepgammatogamma} and ${\cal Q}_{{\boldsymbol{r}} \to \gamma }^{\left( {\xi } \right)}$ in \eqref{mesagertogamma} respectively;
      \State Estimation of the $\xi$-th estimation of $\gamma$ via $\eqref{gammamu}$;
      \State  Updating the messages ${\cal Q}_{{\mathcal{P}_{{\Lambda _0}}} \to {\Lambda _0}}^{\left( {\xi } \right)}$ in \eqref{messageQLambda} and $\mathcal{Q}_{\boldsymbol r \to {\Lambda _0}}^{\left( {\xi } \right)}$ in \eqref{ExmessageQ} respectively;
      \State Estimation of the $\xi$-th estimation of $\Lambda_0$ via $\eqref{LambdaEstimation}$;
      \State Updating the messages ${\cal Q}_{{{\cal P}_{\boldsymbol{x}}} \to {\boldsymbol{x}}}^{\left( {\xi} \right)}$ in \eqref {messageQx} and ${\cal Q}_{{\boldsymbol{r}} \to {\boldsymbol{x}}}^{\left( {\xi} \right)} $ in \eqref{messageQrx} respectively;
     \State Estimation of the $\xi$-th estimation of $\boldsymbol x$ via $\eqref{xithofxmean}$;
      \State ${\boldsymbol x}_i$: Updating the messages $\mathcal{Q}_{{\boldsymbol{r}} \to {{\boldsymbol{x}}_i}}^{\left( \xi  \right)}$ in \eqref {messageQrxi} and ${\cal Q}_{{{\cal P}_{\boldsymbol{x}_i}} \to {\boldsymbol{x}}_i}^{\left( {\xi} \right)} $ in \eqref{messageQxi} respectively;
      \State Estimation of the $\xi$-th estimation of ${\boldsymbol x}_i$ via $\eqref{xthofximean}$;
      \EndWhile
      \State Output the estimation of $\gamma$, $\Lambda_0$, $\boldsymbol x$ and ${\boldsymbol x}_i$
    \label{code:recentEnd}
    \vspace{-0.4em}
  \end{algorithmic}
\end{algorithm}
}

\section{Simulation Results}

\subsection{Simulation  Settings}
In this section, we investigate the estimation performance of the proposed algorithm in different scenarios.  We consider a 2D scenario in the field of $100 m\times100 m$. One target node is localized with the aid of $M=5$ sensor nodes and the parameter parameters are also unknown. The sensor nodes are located coarsely at $\left(10,20\right), \left(80,90\right), \left(30,40\right),\left(10,90\right),\left(60,20\right)$. The location uncertainty of all sensor nodes is set to be $\boldsymbol \Sigma_i = \mu^2\mathbf{I}_2$. The maximum iteration number is $20$ and the threshold is set to be ${\mathcal T}=10^{-3}$.  The true position of the target node is selected randomly in the field with the location uncertainty $\boldsymbol \Sigma  = 100 \mathbf{I}$. The hyper-parameters of prior distributions are given by  ${\mu _\gamma } = 3$, ${\delta _\gamma } = 0.01$ and $\delta _0^2 = \mathbb{E}\left( {\frac{a}{b}} \right) = {10^{ - 6}}$ respectively.

The mentioned settings are unaltered and otherwise stated differently. For comparison, the proposed algorithm is referred to as the JLCE algorithm and compared to the following algorithms: ML-GMP \cite{WCLLiRobust21}, IRGLS \cite{HuangTWC15Source}, and BCRB.
 Given the unknown variable vector ${\boldsymbol{\Theta }}$ and the derivations in \cite{TCRobustLi2020}, we can obtain the total FIM as
\begin{equation}\label{FIM2}
\begin{array}{*{20}{l}}
{\boldsymbol J\left( {\boldsymbol{\Theta }} \right) = \underbrace { - {\mathbb{E}_{{\boldsymbol{r}},{\boldsymbol{\Theta }}}}\left[ {\frac{{{\partial ^2}\ln p\left( {{\boldsymbol{r}}|{\boldsymbol{\Theta }}} \right)}}{{\partial {\boldsymbol{\Theta }}\partial {{\boldsymbol{\Theta }}^T}}}} \right]}_{{\boldsymbol {\cal I}_{\rm{M}}}}\underbrace { - {\mathbb{E}_{\boldsymbol{\Theta }}}\left[ {\frac{{{\partial ^2}\ln p\left( {\boldsymbol{\Theta }} \right)}}{{\partial {\boldsymbol{\Theta }}\partial {{\boldsymbol{\Theta }}^T}}}} \right]}_{{\boldsymbol {\cal I}_{\rm{P}}}},}
\end{array}
\end{equation}
where ${\mathbb E}_{{\boldsymbol  r},\boldsymbol \Theta} \left[ \cdot\right] $ and ${\mathbb E}_{\boldsymbol \Theta} \left[ \cdot\right]$ denote the expectations with respect to the joint distribution $p({\bf r},{\boldsymbol \Theta})$ and prior distribution $p({\boldsymbol \Theta})$, respectively.

The Fisher information matrix in \eqref{FIM2} indicates that the likelihood function and the prior distributions both contribute to the joint estimation of the parameter parameters and target node location.  By splitting the unknown variable ${\boldsymbol{\Theta }}$ into ${{{\boldsymbol{\Theta }}_{\backslash {\boldsymbol{x}}}}}$ and ${\boldsymbol{x}}$, the Fisher information matrix $\boldsymbol J\left( {\boldsymbol{\Theta }} \right)$ can be reformulated to be a symmetric matrix, which is given by
\begin{equation}\label{FIM3}
{\boldsymbol J}\left( {\boldsymbol{\Theta }} \right) = \left[ {\begin{array}{*{20}{c}}
{{{\boldsymbol J}_{\boldsymbol x}}}&{{\boldsymbol J}_{{\boldsymbol x},{{\boldsymbol{\Theta }}_{\backslash {\boldsymbol{x}}}}}^T}\\
{{{\boldsymbol J}_{{\boldsymbol x},{{\boldsymbol{\Theta }}_{\backslash {\boldsymbol{x}}}}}}}&{{{\boldsymbol J}_{{{\boldsymbol{\Theta }}_{\backslash {\boldsymbol{x}}}}}}}
\end{array}} \right],
\vspace{-0.6em}
\end{equation}
where the detailed derivations of the submatrices ${{\boldsymbol J_ {\boldsymbol x}}}$, ${{\boldsymbol J_{\boldsymbol x,{{\boldsymbol{\Theta }}_{\backslash {\boldsymbol{x}}}}}}}$ and ${{\boldsymbol J_{{{\boldsymbol{\Theta }}_{\backslash {\boldsymbol{x}}}}}}}$ are given in Appendix \ref{appendix_alpha}.

Using the Schur complement and equivalent Fisher information matrix in \cite{TCRobustLi2020, LiSecure20IoT},  the $\text{BCRB}$ with respect to ${\boldsymbol x}$ is thus given by
\begin{equation}\label{overallBCRB}
\begin{aligned}
{\rm{BCR}}{{\rm{B}}_{\boldsymbol{x}}}& = {\left[ {{{\boldsymbol J}^{ - 1}}\left( {\boldsymbol{\Theta }} \right)} \right]_{1,1}} + {\left[ {{J^{ - 1}}\left( {\boldsymbol{\Theta }} \right)} \right]_{2,2}} \\& = tr\left(\left( {{{\boldsymbol J}_x} - {\boldsymbol J}_{\boldsymbol x,{{\boldsymbol{\Theta }}_{\backslash {\boldsymbol{x}}}}}^T{\boldsymbol J}_{{{\boldsymbol{\Theta }}_{\backslash {\boldsymbol{x}}}}}^{ - 1}{{\boldsymbol J}_{\boldsymbol x,{{\boldsymbol{\Theta }}_{\backslash {\boldsymbol{x}}}}}}} \right)^{-1}\right).
\end{aligned}
\end{equation}

\subsection{Convergence Properties}
\begin{figure}
  \centering
  \includegraphics[width=2in]{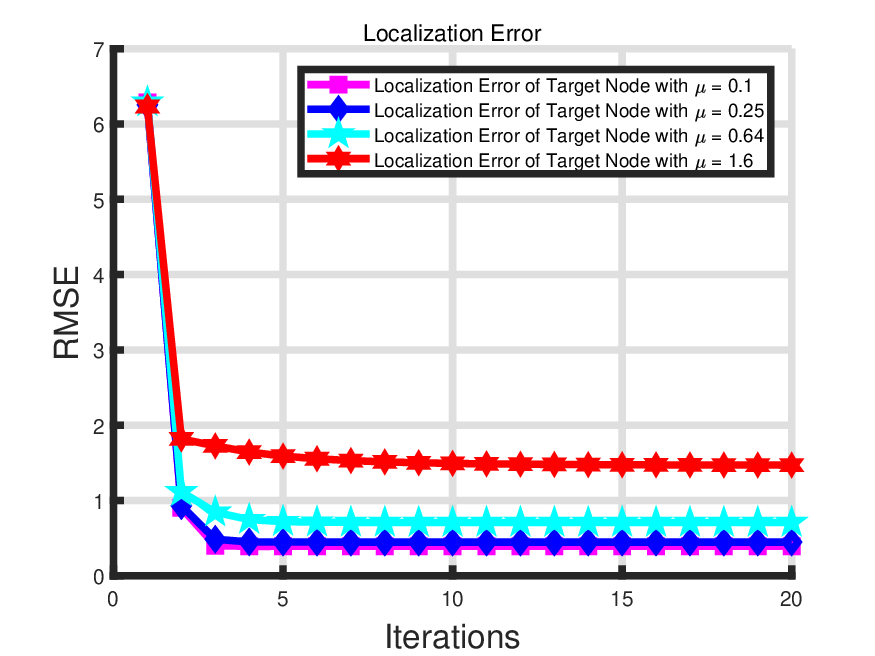}\\
  \caption{The numerical convergence of proposed algorithm on the localization errors of the target node with different sensor node location errors under $10\log\left(\delta_0\right) = -30.$}\label{fig:convergence1}
\end{figure}

\begin{figure}
  \centering
  \includegraphics[width=2in]{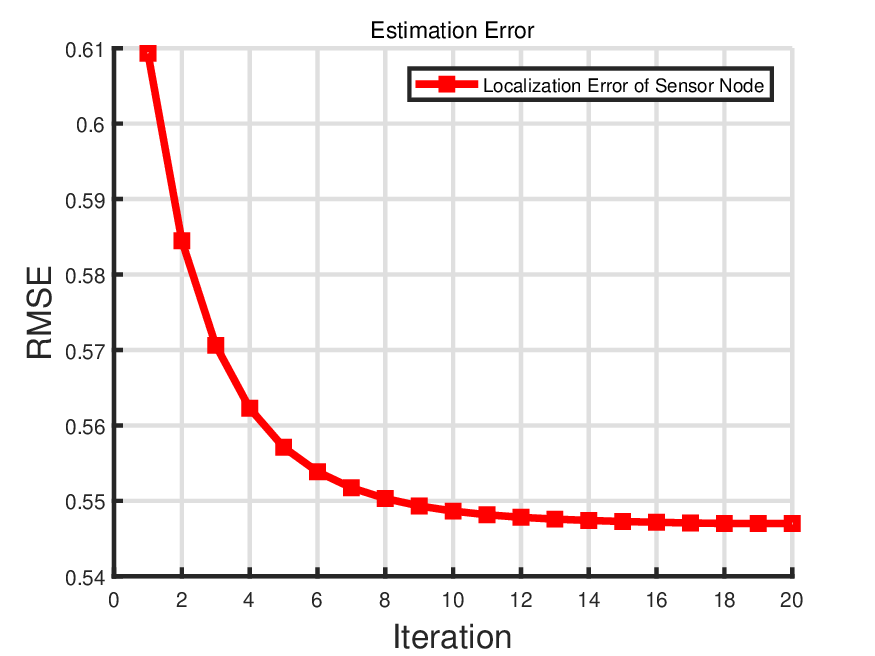}\\
  \caption{The numerical convergence of proposed algorithm on the location refinement of the sensor node under sensor node location errors $\mu = 0.5$ under $10\log\left(\delta_0\right) = -30.$}\label{fig:convergence2}
\end{figure}

\begin{figure}
  \centering
  \includegraphics[width=2in]{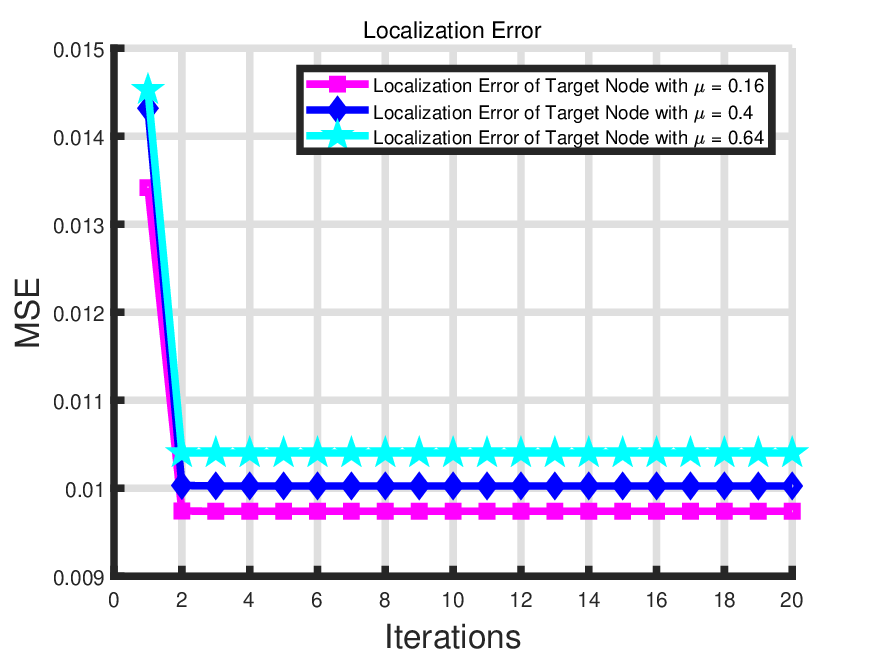}\\
  \caption{The numerical convergence of proposed algorithm on the estimation of $\gamma$ under different sensor node location errors under $10\log\left(\mu\right) = -10$ and $10\log\left(\delta_0\right) = -30.$}\label{fig:convergence3}
\end{figure}

In this subsection, our objective is to investigate the numerical convergence properties of the proposed algorithm across various scenarios. The uncertainties associated with sensor nodes are defined as $\boldsymbol \Sigma_i = \mu^2\mathbf{I}_2$. In Fig.\ref{fig:convergence1}, Fig.\ref{fig:convergence2} and Fig.\ref{fig:convergence3}, we illustrate the localization error of the target node location, the refinement of sensor node locations and the estimation of $\gamma$ respectively. In scrutinizing the numerical convergence properties of the proposed JLCE algorithm in Fig.\ref{fig:convergence1}, Fig.\ref{fig:convergence2}, and Fig.\ref{fig:convergence3}, significant insights are gleaned as follows.

Firstly, the numerical results demonstrate a consistent and monotonic RMSEs or MSEs decrease across successive iterations. These results strongly highlight the remarkable convergence rates achieved by the JLCE algorithm. The localization performance gaps of the target node between different RMSEs straightforwardly indicate the negative impacts of the sensor node location errors over the localization accuracy. Furthermore, the increase in sensor node location errors also are numerically verified to lower the convergence rate of the proposed algorithm.

Moreover, the results illustrated in Fig.\ref{fig:convergence2} strongly highlight refined sensor node locations, signifying a remarkable reduction in location errors. This advantageous outcome is credited to the exchange of positional information orchestrated by the target node, thereby showcasing the collaborative essence of the JLCE algorithm in improving the overall accuracy of node locations. Simultaneously, the distinct convergence rates observed in the proposed algorithm under varying sensor node location errors elucidate a consistent truth that the presence of sensor node location errors diminishes the convergence rate of JLCE. The rationale behind this result lies in the correlation between increased sensor node location errors and heightened location uncertainties, necessitating a more extensive iterative process to mitigate the associated KL divergence.

Additionally, the numerical results in Fig.\ref{fig:convergence2} show the estimation results of the proposed algorithm on the parameter $\gamma$, which not only demonstrate the negative impact of the sensor node location uncertainties on the parameter estimation but also the fast convergence rates of the proposed algorithm.

These numerical findings not only verify the theoretical derivations presented in Section IV but also furnish empirical validation for the gradual enhancement of the variational distribution $q\left(\boldsymbol x\right)$ towards the true posterior distribution $p\left(\boldsymbol x|\boldsymbol r\right)$. Despite uncertainties in the positions of target nodes, the algorithm demonstrates robust performance in both localization and parameter estimation. This resilience against uncertainties underscores the algorithm's effectiveness and reliability in the joint estimation problem. The collective insights drawn from these numerical results emphasize the algorithm's feasibility in addressing the challenges associated with localization and parameter estimation in the distance-dependent noise model.

\subsection{Localization and Estimation Performances}
In this subsection, we dig into the investigation of the localization error and parameter estimation performances of the proposed JLCE algorithm alongside other algorithms through the exploration of various examples. This comparative investigation seeks to verify the effectiveness of the JLCE algorithm in the presence of uncertainties associated with sensor node positions and provides valuable insights into its performance compared to other methodologies. In this analysis, we assume that the expectations of parameters are known for all algorithms under consideration.  It is crucial to highlight that sensor node location errors are taken into account for all algorithms studied, supporting a comprehensive evaluation of their robustness in real physical scenarios.  Through this rigorous analysis, we aim to perceive the strengths and limitations of the JLCE algorithm, shedding light on its applicability and superiority in addressing challenges related to joint localization and parameter estimation.

\begin{figure}
  \centering
  \includegraphics[width=2in]{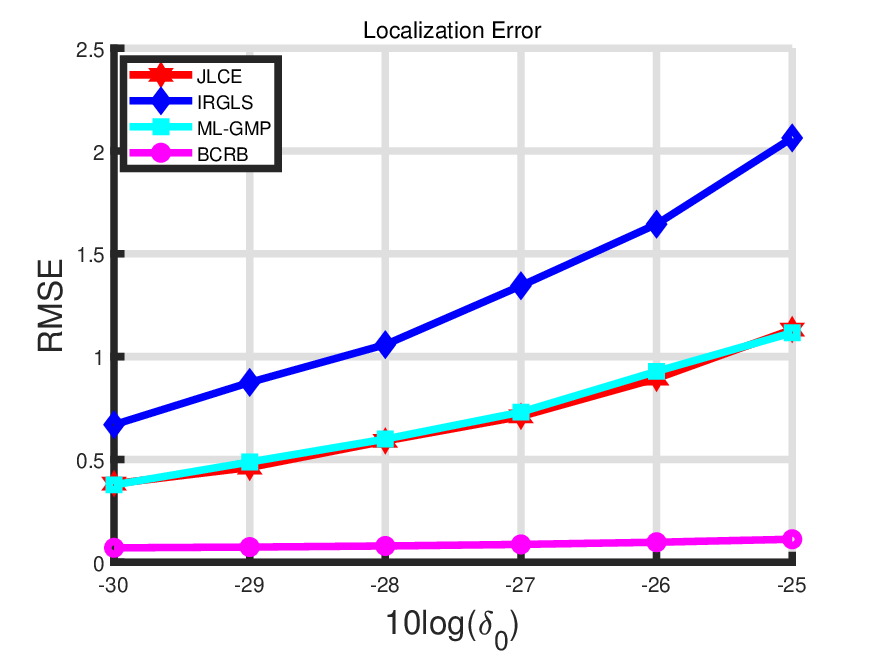}\\
  \caption{The localization errors of proposed algorithm JLCE and other algorithms with different reference noise power $\delta_0$  under $10\log\left(\mu\right) = -10$ and $10\log\left(\delta_0\right) = -30.$}\label{fig:snr}
\end{figure}

In Fig.\ref{fig:snr}, we dig into the investigation of the localization performance of the proposed JLCE algorithm and other algorithms across varying reference noise powers $\delta_0$. The superior performance of the JLCE algorithm outperforms the IRGLS algorithm, which can be attributed to the incorporation of prior information and the MMSE criterion. Moreover, the comparable accuracy achieved by JLCE and the ML-GMP algorithm stems from their common utilization of posterior distributions for both sensor nodes and the target node in the localization process. The numerical results unequivocally indicate the superior effectiveness of the proposed JLCE algorithm.

\begin{figure}
  \centering
  \includegraphics[width=2in]{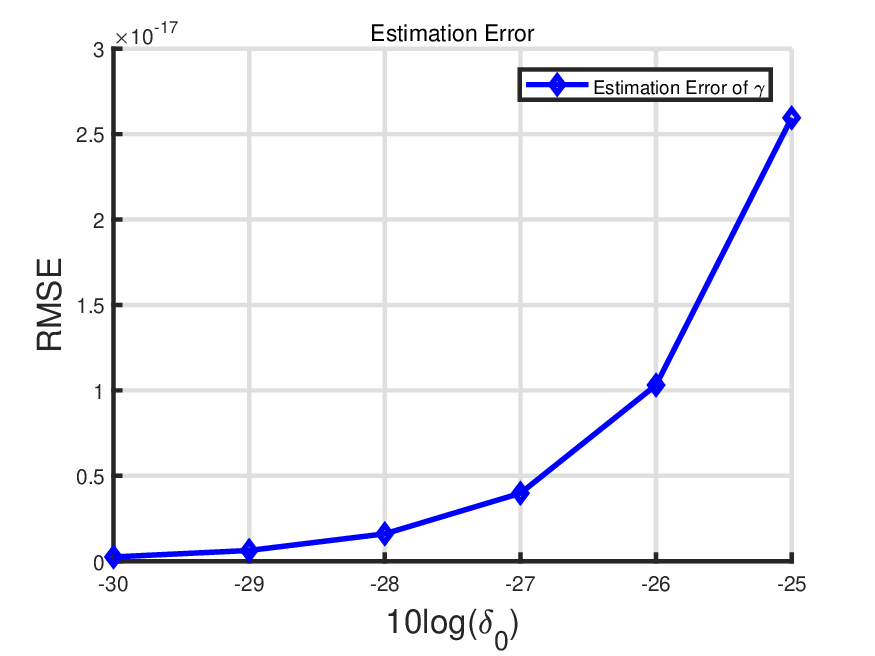}\\
  \caption{The estimation errors of the path loss exponent $\gamma$ of proposed algorithm JLCE with different reference noise power $\delta_0$ under $10\log\left(\mu\right) = -10.$}\label{fig:gammasnr}
\end{figure}

It is important to note that the benchmark BCRB is directly affected by noise power, comprising the reference noise power $\delta_0$ and the unknown distance $d_i$. As such, the rapid variations in reference noise power have a marginal impact on the benchmark, resulting in slight increases in the BCRB. In Fig.\ref{fig:gammasnr}, we present the estimation error of parameter parameters, $\gamma$ achieved by the JLCE algorithm. The numerical results underscore the algorithm's ability to attain accurate parameter parameter estimations. Intuitively, as the reference noise power increases, the estimation errors also escalate, correlating with the observed rise in localization errors depicted in Fig.\ref{fig:snr}. This is inherently correlated to the interplay between reference noise power and localization accuracy, indicating the intricate relationship between parameter estimation and noise model.


\begin{figure}
  \centering
  \includegraphics[width=2in]{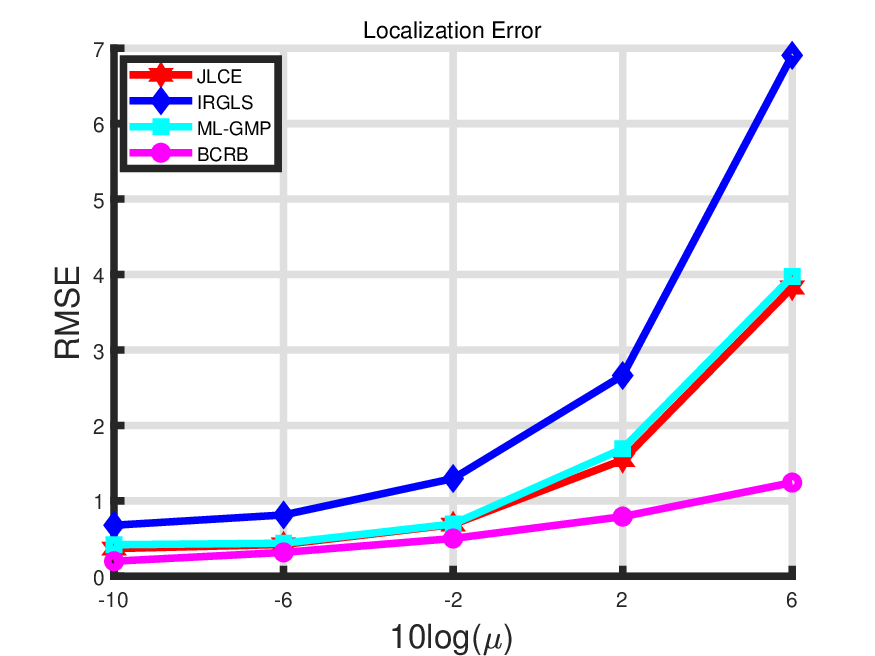}\\
  \caption{The localization errors of proposed algorithm JLCE and other algorithms with different sensor node uncertainties under $10\log\left(\delta_0\right) = -30$.}\label{fig:UI}
\end{figure}

\begin{figure}
  \centering
  \includegraphics[width=2in]{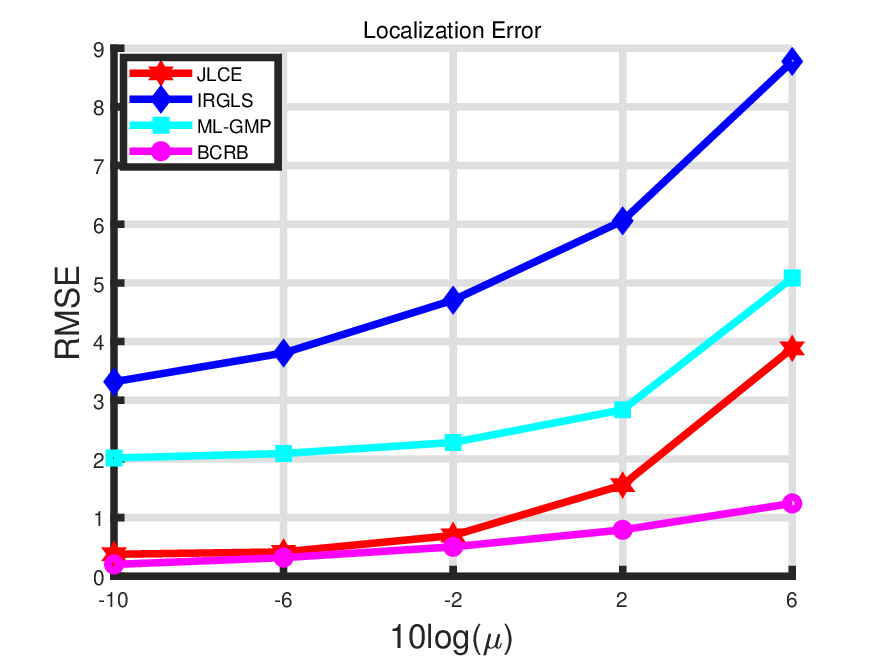}\\
  \caption{The localization errors of proposed algorithm JLCE and other algorithms with different sensor node uncertainties $\mu$ under $\boldsymbol \Delta_s = 50$ and $10\log\left(\delta_0\right) = -30$.}\label{fig:UI_far}
\end{figure}

\begin{figure}
  \centering
  \includegraphics[width=2in]{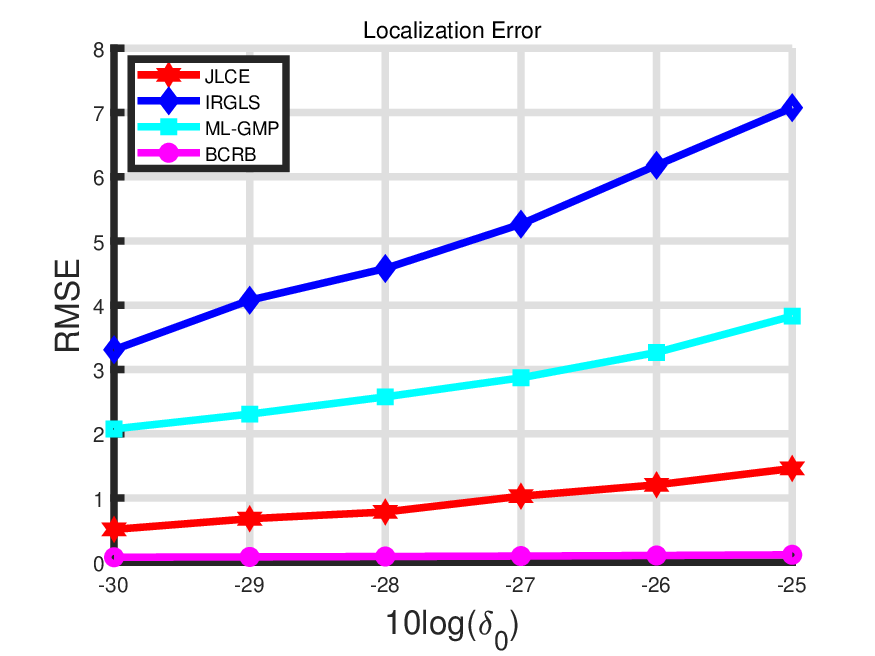}\\
  \caption{The localization errors of proposed algorithm JLCE and other algorithms with different reference noise power $\delta_0$ under $\boldsymbol \Delta_s = 50$ and $10\log\left(\mu\right) = -10.$}\label{fig:snr_far}
\end{figure}

\begin{figure}
  \centering
  \includegraphics[width=2in]{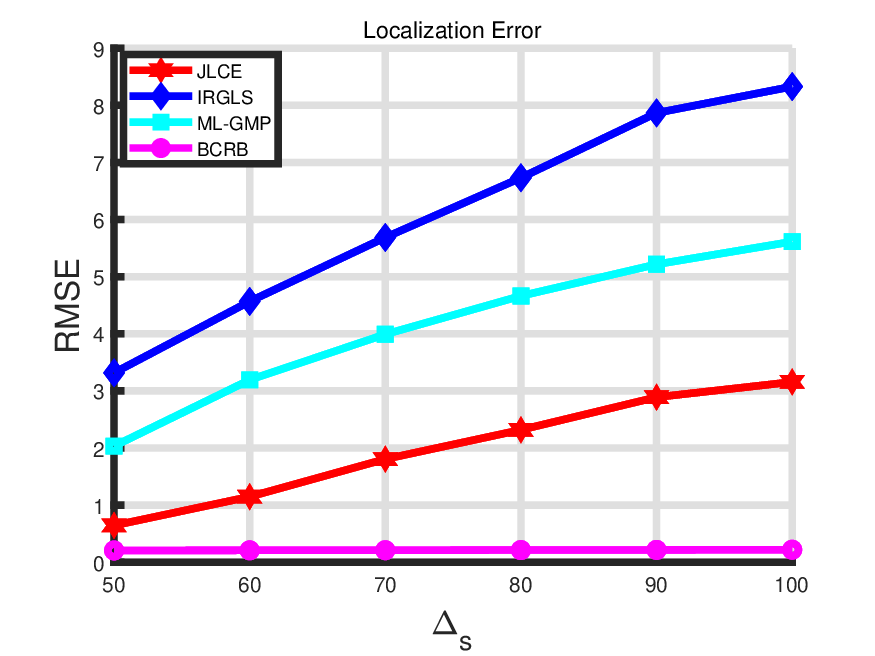}\\
  \caption{The localization errors of proposed algorithm JLCE and other algorithms with different sensor location offsets $\boldsymbol \Delta_s$ under $10\log\left(\delta_0\right) = -30$ under $10\log\left(\mu\right) = -10.$}\label{fig:delta_far}
\end{figure}

In Fig.\ref{fig:UI}, we focus on the investigation of the localization performance of the proposed JLCE algorithm, along with other algorithms, across varying levels of sensor node uncertainties $\mu_i$. The numerical findings compellingly illustrate the adverse effects of sensor node location errors on the localization performance of the target node. Moreover, the JLCE algorithm outperforms the IRGLS algorithm, showcasing its robustness in mitigating the negative impact of sensor node uncertainties. Furthermore, the comparable localization performances of the JLCE and ML-GMP algorithms align with expectations, as both leverage the MMSE criterion and incorporate prior distributions of sensor nodes and the target node in the localization process. These results are consistent with the findings reported in \cite{TCRobustLi2020}, highlighting the significant impact of sensor node location errors on the localization benchmark, a trend corroborated by the outcomes presented in Fig.\ref{fig:UI}. The superiority of the proposed JLCE algorithm in navigating and minimizing the impact of sensor node uncertainties positions it as a promising solution for robust localization in real-world scenarios.
To further investigate the impacts of distance over the estimation and localization performances, we enlarge the distances between the sensor nodes and target nodes by adding a coordinate offset $\boldsymbol \Delta_s$ to the sensor node locations. In Fig.\ref{fig:UI_far} and Fig.\ref{fig:snr_far}, the coordinate offset $\boldsymbol \Delta_s$ is set to be ${{\boldsymbol{\Delta }}_s} = \left[ {50,50} \right]$ and the localization performances are investigated under different sensor node location errors $\mu$ and different different reference noise powers $\delta_0$. The simulation results both indicate the proposed algorithm can achieve superior performance than the other algorithms under greater distance-dependent noise.  Comparing the results in Fig.\ref{fig:UI} and Fig.\ref{fig:UI_far},  it is straightforward that greater distances bring negative impacts on the localization performances, which is validated in Fig.\ref{fig:delta_far}.  The ML-GMP algorithm is much more sensitive to the location offsets between the sensors and the target node, which results from the ML approximation step of ignoring the distance-dependent noise. The proposed algorithm merely are affected by the location offsets due to the adoption of Taylor approximation over the complicated noise. These three results demonstrate that the proposed algorithm can achieve robust performances in worse noise environments.

\section{Conclusion}
The paper addresses a significant challenge in the field by exploring a joint problem involving localization and parameter estimation, taking into account both distance-dependent noise and uncertainties associated with sensor node positions. To tackle the inherent nonlinearity and non-convexity of the objective distribution function, a novel message-passing approach is proposed. This method leverages a message-passing algorithm to approximate the true posterior distribution and employs Taylor expansion to linearize the objective function. Subsequently, the target node location and parameters are estimated using either the MMSE or MAP techniques. Additionally, the paper digs into an investigation of the complexity and convergence characteristics of the proposed algorithm. The simulation results presented in the paper demonstrate the superior performance of the proposed algorithm in terms of parameter estimation and localization accuracy across various illustrative examples. This innovative contribution not only addresses the intricacies of joint localization and parameter estimation but also presents the practical effectiveness of the proposed algorithm through rigorous simulations.

\appendices
\section{}\label{appendix_alpha}

According to the equation \eqref{FIM2}, the submatrice ${{\boldsymbol J_ {\boldsymbol x}}}$ can be given by
\begin{equation}\label{FIMJx}
\begin{array}{*{20}{l}}
{{\boldsymbol J_{\boldsymbol{x}}} = \underbrace { - {\mathbb{E}_{{\boldsymbol{r}},{\boldsymbol{\Theta }}}}\left[ {\frac{{{\partial ^2}\ln p\left( {{\boldsymbol{r}}|{\boldsymbol{\Theta }}} \right)}}{{\partial {\boldsymbol{x}}\partial {{\boldsymbol{x}}^T}}}} \right]}_{{\boldsymbol {\cal I}^{{\boldsymbol{x}},{\boldsymbol{x}}}_{\rm{M}}}}\underbrace { - {\mathbb{E}_{\boldsymbol{\Theta }}}\left[ {\frac{{{\partial ^2}\ln p\left( {\boldsymbol{x}} \right)}}{{\partial {\boldsymbol{x}}\partial {{\boldsymbol{x}}^T}}}} \right]}_{{\boldsymbol {\cal I}^{{\boldsymbol{x}},{\boldsymbol{x}}}_{\rm{P}}}},}
\end{array}
\end{equation}

By substituting the likelihood function $\eqref{likelihood}$ and prior distributions $\eqref{priorX}$ into \eqref{FIMJx}, it yields
\begin{equation}\label{Ipxx}
\boldsymbol {\cal I}_{\rm P}^{{\boldsymbol{x}},{\boldsymbol{x}}} =  - {\mathbb{E}_{\boldsymbol{\Theta }}}\left[ {\frac{{{\partial ^2}\ln p\left( {\boldsymbol{x}} \right)}}{{\partial {\boldsymbol{x}}\partial {{\boldsymbol{x}}^T}}}} \right] = {{\boldsymbol{\Sigma }}^{ - 1}},
\end{equation}
\begin{equation}\label{Imxx}
\begin{array}{l}
\boldsymbol {\cal I}_{\rm M}^{{\boldsymbol{x}},{\boldsymbol{x}}} =  - {\mathbb{E}_{{\boldsymbol{r}},{\boldsymbol{\Theta }}}}\left[ {\frac{{{\partial ^2}\ln p\left( {{\boldsymbol{r}}|{\boldsymbol{\Theta }}} \right)}}{{\partial {\boldsymbol{x}}\partial {{\boldsymbol{x}}^T}}}} \right]=  {\mathbb{E}_{{\boldsymbol{r}},{\boldsymbol{\Theta }}}}\left[ {\sum\limits_{i = 1}^N {\frac{{{\partial ^2}h\left( {{\boldsymbol{x}},{{\boldsymbol{x}}_i}} \right)}}{{\partial {\boldsymbol{x}}\partial {{\boldsymbol{x}}^T}}}} } \right].
\end{array}
\end{equation}
where $h\left( {{\boldsymbol{x}},{{\boldsymbol{x}}_i}} \right) =\sum\limits_{k=1}^{2} { h_k\left( {{\boldsymbol{x}},{{\boldsymbol{x}}_i}} \right)}$ .

The above expectations involve nonlinear expectations with respect to the complicated distributions and it is extremely challenging to find the closed-form expressions. To  obtain the approximation to the estimation performance benchmark, we apply second-order Taylor expansion to the nonlinear expectation ${\mathbb{E}_p}\left( {{ \mathcal{Z}}\left( \zeta  \right)} \right)$ as follows
\begin{equation}\label{exTaylor}
\begin{aligned}
{\mathbb{E}_p}\left( {Z\left( \zeta  \right)} \right)& \approx {\mathbb{E}_p}\left( {Z\left( {{\mu _\zeta }} \right)} \right) + {\mathbb{E}_p}\left( {{{\left. {\frac{{\partial Z}}{{\partial \zeta }}} \right|}_{\xi  = {\mu _\zeta }}}\left( {\zeta  - {\mu _\zeta }} \right)} \right) \\ &+ {\mathbb{E}_p}\left( {\frac{1}{2}{{\left. {\frac{{{\partial ^2}\mathcal{Z}}}{{\partial {\zeta ^2}}}} \right|}_{\xi  = {\mu _\zeta }}}{{\left( {\zeta  - {\mu _\zeta }} \right)}^2}} \right)\\
 & \approx {\mathbb{E}_p}\left( {\mathcal{Z}\left( {{\mu _\zeta }} \right)} \right) +\frac{1}{2}{{\left. {\frac{{{\partial ^2}\mathcal{Z}}}{{\partial {\zeta ^2}}}} \right|}_{\xi  = {\mu _\zeta }}}{\delta _\zeta },
\end{aligned}
\end{equation}
where ${\mathbb{E}_p}$ takes expectation with respect to $p\left(\zeta\right)$. ${\mu _\zeta }$ and $\delta_{\zeta}$ are the mean and variance of the distribution  $p\left(\zeta\right)$. Meanwhile, the approximation holds with continuous and sufficiently differentiable ${{ \mathcal{Z}}\left( \zeta  \right)}$. Therefore, the expectation in \eqref{Imxx} can be obtained by replacing the corresponding terms with tedious expectation approximations.

According to the equation \eqref{FIM2}, the submatrices ${{\boldsymbol J_{\boldsymbol x,{{\boldsymbol{\Theta }}_{\backslash {\boldsymbol{x}}}}}}}$ can be given by
\begin{equation}\label{FIMxthetax}
{\boldsymbol J_{{\boldsymbol{x}},{{\boldsymbol{\Theta }}_{\backslash {\boldsymbol{x}}}}}} = \underbrace { - {\mathbb{E}_{{\boldsymbol{r}},{\boldsymbol{\Theta }}}}\left[ {\frac{{{\partial ^2}\ln p\left( {{\boldsymbol{r}}|{\boldsymbol{\Theta }}} \right)}}{{\partial {\boldsymbol{x}}\partial {\boldsymbol{\Theta }}_{\backslash {\boldsymbol{x}}}^T}}} \right]}_{\boldsymbol {\cal I}_{\rm M}^{{\boldsymbol{x}},{{\boldsymbol{\Theta }}_{\backslash {\boldsymbol{x}}}}}},
\end{equation}
which includes the following components
\begin{equation}\label{FIM}
{{\boldsymbol{J}}_{{\boldsymbol{x}},{{\boldsymbol{\Theta }}_{\backslash {\boldsymbol{x}}}}}} = \left[ {\begin{array}{*{20}{c}}
{{{\boldsymbol{J}}_{{\boldsymbol{x}},{{\boldsymbol{x}}_1}}}}& \ldots &{{{\boldsymbol{J}}_{{\boldsymbol{x}},{{\boldsymbol{x}}_N}}}}&{{{\boldsymbol{J}}_{{\boldsymbol{x}},\gamma }}}&{{{\boldsymbol{J}}_{{\boldsymbol{x}},{\Lambda _0}}}}
\end{array}} \right].
\end{equation}

Specifically, these components can be calculated respectively
\begin{equation}\label{com1}
\begin{aligned}
{{\boldsymbol{J}}_{{\boldsymbol{x}},{{\boldsymbol{x}}_i}}} & =  - {\mathbb{E}_{{\boldsymbol{r}},{\boldsymbol{\Theta }}}}\left( {\frac{{\partial \ln p\left( {{\boldsymbol{r}}\mid {\boldsymbol{\Theta }}} \right)}}{{\partial {\boldsymbol{x}}\partial {\boldsymbol{x}}_i^T}}} \right) \\ &= {\mathbb{E}_{\boldsymbol{\Theta }}}\left( {\frac{{2{\Lambda _0} - {\gamma ^2}d_i^{\gamma  - 2}}}{{2d_i^\gamma }}\frac{{\partial {d_i}}}{{\partial {\boldsymbol{x}}}}\frac{{\partial {d_i}}}{{\partial {\boldsymbol{x}}_i^T}}} \right),
\end{aligned}
\end{equation}

\begin{equation}\label{com2}
\begin{aligned}
{{\boldsymbol{J}}_{{\boldsymbol{x}},\gamma }} & =  - {\mathbb{E}_{{\boldsymbol{r}},{\boldsymbol{\Theta }}}}\left( {\frac{{\partial \ln p\left( {{\boldsymbol{r}}\mid {\boldsymbol{\Theta }}} \right)}}{{\partial {\boldsymbol{x}}\partial {\gamma ^T}}}} \right) \\ & = {\mathbb{E}_{\boldsymbol{\Theta }}}\left( {\frac{1}{2}\frac{1}{{{d_i}}}\frac{{\partial {d_i}}}{{\partial {\boldsymbol{x}}}} - \frac{{d_i^{ - 1} - \gamma d_i^{ - 1}\ln {d_i}}}{2}\frac{{\partial {d_i}}}{{\partial {\boldsymbol{x}}}}} \right),
\end{aligned}
\end{equation}
and
\begin{equation}\label{com3}
{{\boldsymbol{J}}_{{\boldsymbol{x}},{\Lambda _0}}} =  - {\mathbb{E}_{{\boldsymbol{r}},{\boldsymbol{\Theta }}}}\left( {\frac{{\partial \ln p\left( {{\boldsymbol{r}}\mid {\boldsymbol{\Theta }}} \right)}}{{\partial {\boldsymbol{x}}\partial \Lambda _o^T}}} \right) = {\mathbb{E}_{\boldsymbol{\Theta }}}\left( {\frac{{\gamma d_i^{ - 1}}}{{2{\Lambda _0}}}\frac{{\partial {d_i}}}{{\partial {\boldsymbol{x}}}}} \right),
\end{equation}
where expectations can be obtained via the expectation approximations in \eqref{exTaylor}.

According to the equation \eqref{FIM2}, the submatrice ${{\boldsymbol{J}}_{{{\boldsymbol{\Theta }}_{\backslash {\boldsymbol{x}}}},{{\boldsymbol{\Theta }}_{\backslash {\boldsymbol{x}}}}}} $ can be given by
\begin{equation}\label{FIMtheta}
{{\boldsymbol{J}}_{{{\boldsymbol{\Theta }}_{\backslash {\boldsymbol{x}}}},{{\boldsymbol{\Theta }}_{\backslash {\boldsymbol{x}}}}}} = \underbrace { - {\mathbb{E}_{{\boldsymbol{r}},{\boldsymbol{\Theta }}}}\left[ {\frac{{{\partial ^2}\ln p\left( {{\boldsymbol{r}}|{\boldsymbol{\Theta }}} \right)}}{{\partial {{\boldsymbol{\Theta }}_{\backslash {\boldsymbol{x}}}}\partial {\boldsymbol{\Theta }}_{\backslash {\boldsymbol{x}}}^T}}} \right]}_{\boldsymbol {\cal I}_{\rm{M}}^{{{\boldsymbol{\Theta }}_{\backslash {\boldsymbol{x}}}},{{\boldsymbol{\Theta }}_{\backslash {\boldsymbol{x}}}}}}\underbrace { - {\mathbb{E}_{\boldsymbol{\Theta }}}\left[ {\frac{{{\partial ^2}\ln p\left( {\boldsymbol{x}} \right)}}{{\partial {{\boldsymbol{\Theta }}_{\backslash {\boldsymbol{x}}}}\partial {\boldsymbol{\Theta }}_{\backslash {\boldsymbol{x}}}^T}}} \right]}_{{\boldsymbol{\cal I}}_{\rm{P}}^{{{\boldsymbol{\Theta }}_{\backslash {\boldsymbol{x}}}},{{\boldsymbol{\Theta }}_{\backslash {\boldsymbol{x}}}}}}.
\end{equation}

By substituting the likelihood function $\eqref{likelihood}$ and prior distributions $\eqref{messQgamma}$, $\eqref{messQLambda}$ and $\eqref{priorXi}$ into \eqref{FIMJx}, it yields
\begin{equation}\label{excom1}
\begin{aligned}
{{\boldsymbol{J}}_{{{\boldsymbol{x}}_i},{{\boldsymbol{x}}_i}}} &=  - {\mathbb{E}_{{\boldsymbol{r}},{\boldsymbol{\Theta }}}}\left( {\frac{{{\partial ^2}\ln p\left( {{\boldsymbol{r}}|{\boldsymbol{\Theta }}} \right)}}{{\partial {{\boldsymbol{x}}_i}\partial {\boldsymbol{x}}_i^T}}} \right) - {\mathbb{E}_{{\boldsymbol{r}},{\boldsymbol{\Theta }}}}\left( {\frac{{{\partial ^2}\ln p\left( {{{\boldsymbol{x}}_i}} \right)}}{{\partial {{\boldsymbol{x}}_i}\partial {\boldsymbol{x}}_i^T}}} \right)\\
 &=  {\mathbb{E}_{{\boldsymbol{r}},{\boldsymbol{\Theta }}}}\left[ {\frac{{{\partial ^2}h\left( {{\boldsymbol{x}},{{\boldsymbol{x}}_i}} \right)}}{{\partial {\boldsymbol{x}}\partial {{\boldsymbol{x}}^T}}}} \right] + {\boldsymbol{\Sigma }}_i^{ - 1},
\end{aligned}
\end{equation}
\begin{equation}\label{excom2}
\begin{aligned}
{{\boldsymbol{J}}_{\gamma ,\gamma }} &=  - {\mathbb{E}_{{\boldsymbol{r}},{\boldsymbol{\Theta }}}}\left( {\frac{{{\partial ^2}\ln p\left( {{\boldsymbol{r}}|{\boldsymbol{\Theta }}} \right)}}{{\partial \gamma \partial {\gamma ^T}}}} \right) - {\mathbb{E}_{\boldsymbol{\Theta }}}\left( {\frac{{{\partial ^2}\ln p\left( \gamma  \right)}}{{\partial \gamma \partial {\gamma ^T}}}} \right)\\
 & = {\mathbb{E}_{\boldsymbol{\Theta }}}\left( {\frac{1}{2}{{\left( {\ln {d_i}} \right)}^2}} \right) + \delta _\gamma ^{ - 1},
\end{aligned}
\end{equation}
\begin{equation}\label{excom3}
\begin{aligned}
{{\boldsymbol{J}}_{{\Lambda _0},{\Lambda _0}}} & =  - {\mathbb{E}_{{\boldsymbol{r}},{\boldsymbol{\Theta }}}}\left( {\frac{{{\partial ^2}\ln p\left( {{\boldsymbol{r}}|{\boldsymbol{\Theta }}} \right)}}{{\partial {\Lambda _0}\partial \Lambda _0^T}}} \right) - {\mathbb{E}_{\boldsymbol{\Theta }}}\left( {\frac{{{\partial ^2}\ln p\left( {{\Lambda _0}} \right)}}{{\partial {\Lambda _0}\partial \Lambda _0^T}}} \right) \\& = {\mathbb{E}_{\boldsymbol{\Theta }}}\left( {\left( {0.5 + b} \right)\Lambda _0^{ - 2}} \right).
\end{aligned}
\end{equation}

\ifCLASSOPTIONcaptionsoff
  \newpage
\fi

\bibliographystyle{IEEEtran}
\bibliography{distancedependent}

\end{document}